\documentclass[11pt]{article}

\usepackage[latin1]{inputenc}
\usepackage{graphicx}
\usepackage{color}
\usepackage{graphicx}
\usepackage{comment}
\usepackage[mathscr]{euscript}
\usepackage{algorithm}
\usepackage[noend]{algpseudocode}
\usepackage{enumitem}
\usepackage{lscape}
\usepackage{makecell}
\usepackage{lscape}
\usepackage{verbatim}
\usepackage[colorlinks=true,  linkcolor=blue, citecolor=blue, urlcolor=blue]{hyperref}
\usepackage{booktabs, multirow}
\usepackage{amssymb,amsmath}
\usepackage{float}
\usepackage{xcolor}
\usepackage[round]{natbib}

\newcommand{\aalpha}{\mbox{\boldmath{$\alpha$}}}
\newcommand{\ppsi}{\mbox{\boldmath{$\psi$}}}

\newcommand{\ttheta}{\mbox{\boldmath{$\theta$}}}
\newcommand{\ssigma}{\mbox{\boldmath{$\sigma$}}}
\newcommand{\rrho}{\mbox{\boldmath{$\rho$}}}

\newcommand{\Ttheta}{\mbox{\boldmath{$\Theta$}}}
\newcommand{\Pphi}{\mbox{\boldmath{$\Phi$}}}
\newcommand{\Oomega}{\mbox{\boldmath{$\Omega$}}}

\begin{document}


\title{\textbf{High-dimensional order-free multivariate spatial disease mapping}}

\author{Vicente, G.$^{1,2,3}$, Adin, A.$^{2,3,4}$, Goicoa, T.$^{2,3,4}$ and Ugarte, M.D.$^{2,3,4}$\\
\small {\textit{$^1$ Facultad de Ciencias Econ{\'o}micas, Universidad de Cuyo, Argentina.}} \\
\small {\textit{$^2$ Department of Statistics, Computer Sciences and Mathematics, Public University of Navarre, Spain.}} \\
\small {\textit{$^3$ Institute for Advanced Materials and Mathematics, InaMat$^2$, Public University of Navarre, Spain.}}\\
\small {\textit{$^4$ IdiSNA, Health Research Institute of Navarre, Recinto de Complejo Hospitalario de Navarra, Spain.}}\\
\small {$*$Correspondence to María Dolores Ugarte, Departamento de Estad\'istica, Inform\'atica y Matem\'aticas, } \\
\small {Universidad P\'ublica de Navarra, Campus de Arrosadia, 31006 Pamplona, Spain.} \\
\small {\textbf{E-mail}: lola@unavarra.es }}
\date{}

\makeatletter
\pdfbookmark[0]{\@title}{title}
\makeatother

\maketitle

\begin{abstract}
Despite the amount of research on disease mapping in recent years, the use of multivariate models for areal spatial data remains limited due to difficulties in implementation and computational burden. These problems are exacerbated when the number of small areas is very large. In this paper, we introduce an order-free multivariate scalable Bayesian modelling approach to smooth mortality (or incidence) risks of several diseases simultaneously. The proposal partitions the spatial domain into smaller subregions, fits multivariate models in each subdivision and obtains the posterior distribution of the relative risks across the entire spatial domain. The approach also provides posterior correlations among the spatial patterns of the diseases in each partition that are combined through a consensus Monte Carlo algorithm to obtain correlations for the whole study region. We implement the proposal using integrated nested Laplace approximations (INLA) in the R package \texttt{bigDM} and use it to jointly analyse colorectal, lung, and stomach cancer mortality data in Spanish municipalities. The new proposal permits the analysis of big data sets and provides better results than fitting a single multivariate model.
\end{abstract}


\bigskip

\section{Introduction} \label{sec:introduction}
Research on methodology for the spatial (and spatio-temporal) analysis of areal count data has grown tremendously in the last years, and statistical models have proven an essential tool for studying the geographic distribution of data in small areas.
The main objective of these techniques is to smooth standardized mortality (incidence) ratios or crude rates to discover geographic patterns of the phenomenon under study.
These models and methods have been mainly applied in epidemiology to analyse the incidence and mortality of chronic diseases such as cancer, but some recent research has demonstrated their applicability to the spatial and spatio-temporal analysis of crimes \citep[see for example][]{LiRich2014}, and in particular crimes against women \citep[see for example][]{vicente2018small,vicente2020dowry}. Although research on single disease analysis has been very fruitful and abundant since the seminal work of \cite{besag1991bayesian}, joint modelling of several responses offers some advantages. The first is that it improves smoothing by borrowing strength between diseases. The second, and probably more important, is that it allows to establish relationships between different diseases in terms of similar or completely different geographical distributions, i.e. in terms of correlations between spatial patterns.
This is crucial, as these correlations may indicate associations with common underlying risk factors and certain (usually unknown) connections between the different diseases. The joint analysis is carried out through multivariate spatial models that can cope with both spatial correlation within diseases and correlation between diseases. Not only can multivariate models account for correlation between diseases, but also improve estimates by borrowing information from nearby areas.

\noindent There is a considerable amount of research on Bayesian multivariate spatial models for count data, most of the proposals relying on Markov chain Monte Carlo (MCMC) algorithms for estimation and inference. However, their use in practice is still limited due to a lack of \lq\lq easy to use'' implementations of the models in statistical packages and the computational burden of most of the proposals that preclude practitioners from exploiting their undoubted advantages over univariate counterparts.
A comprehensive review of the subject can be found in the work of \cite{MacNab2018Test} which discusses the three main lines in the construction of multivariate proposals based on Gaussian Markov random fields. Namely, a multivariate conditionals-based approach \citep{mardia1988multi}, a univariate conditionals-based approach \citep{sain2011}, and a coregionalization framework \citep{jin2007order}. Regarding the latter, \cite{martinez2013general} derives a general theoretical setting for multivariate areal models that covers many of the existing proposals in the literature. However, this procedure is unaffordable for a moderate to large number of diseases due to the high computational cost of the MCMC algorithms. \cite{botella2015unifying} reformulate the Mart{\'i}nez-Beneito framework and present the so called M-models as a simpler and more computationally efficient alternative.
This approach makes it possible to increase the number of diseases in the model at the expense of the identifiability of certain parameters.
Recently, \cite{vicente2020SERRA} consider the M-models-based approach to analyse in space and time different crimes against women in India. These authors estimate the M-models using integrated nested Laplace approximations (INLA) and numerical integration for Bayesian inference  \citep[see][]{rue2009approximate} and implement the procedure using the \textsf{'rgeneric'} construction in R-INLA \citep{LindRue2015}. The result is a \lq\lq ready-to-use'' function  for a wide audience with limited programming skills.

\noindent Several alternatives to Gaussian Markov random fields have been also proposed in the disease mapping literature.  A very attractive modelling approach is the use of splines to smooth risks \citep{goicoa2012comparing}. Research on multivariate spline models for fitting spatio-temporal count data is not so abundant and focuses on multivariate structures to deal with the spatial and temporal dependence for one response measured in several time periods \citep[see for example][]{macnab2016lineara,ugarte2010spatio,ugarte2017one}.
Very recently, \cite{vicente2021Psplines} propose multivariate P-spline models to study the spatio-temporal evolution of four crimes against women. 
Unfortunately, inference for these multivariate proposals (and also for univariate approaches) become unfeasible when the number of areas is very large, and the scalability of the procedures is an issue.

\noindent New directions in disease mapping points towards developing new methods for Bayesian inference when the number of small areas is very large \citep[][]{MacNab2022SpatStat}. Creating computationally efficient methods for large data sets is one of the greatest challenges in the field of univariate and multivariate spatial statistics. Several methods for massive geostatistical data (point-referenced) have been already proposed \cite[see for example][among others]{cressie2008fixed, lindgren2011explicit, nychka2015multiresolution, katzfuss2017Mult, katzfuss2021general}.
However, in the case of areal (lattice) count data, research on the scalability of statistical models is not so abundant.
Recently, \cite{orozco2021scalable,orozco2022} propose a scalable Bayesian modelling approach for univariate high-dimensional spatial and spatio-temporal disease mapping data. They propose to divide the spatial domain into \textit{D} subregions where independent models can be fitted simultaneously. To avoid the border effect in the risk estimates, \textit{k}-order neighbours are added to each subregion so that some areal units will have several risk estimates. Finally, a unique posterior distribution for these risks is obtained by either computing the mixture distribution of the estimated posterior probability density functions or by selecting the posterior marginal risk estimate corresponding to the original domain to which the area belongs. This proposal reduces computational time and, in contrast to fitting a single model to the whole domain, it allows different degree of spatial smoothness over the areas within the different subdomains.

\noindent The main objective of this paper is to present a new approach to fit order-free multivariate spatial disease mapping models in domains with a very large number of small areas avoiding high RAM/CPU usage, and making it accessible to users with limited computing facilities.
In particular, we combine the \cite{orozco2021scalable,orozco2022} \lq\lq divide-and-conquer'' approach with a modification of the \cite{botella2015unifying} M-models to avoid overparametrization. 
An additional interesting novelty of our proposal is that we are able to retrieve both the posterior distributions of the correlations between the spatial patterns of each disease in the whole spatial domain, as well as in each of the subdivisions. We have implemented the methodology in INLA to reduce computational burden through our R package \texttt{bigDM}  \citep{bigDM}, that also implements recent high-dimensional univariate proposals.


\noindent The rest of the article has the following structure. Section~\ref{sec:Mmodels} reviews the M-models to fit multivariate data. In Section~\ref{sec:high_dim} we present the new methodology to make the multivariate models scalable. In Section~\ref{sec:simulation}, we conduct a simulation study to compare the performance of this new modelling approach with a single multivariate spatial M-model fitted to the whole domain. Finally, in Section~\ref{sec:Case_study}, we use the new proposal to jointly analyse lung, colorectal and stomach cancer male mortality in Spanish municipalities. The paper closes with a discussion.

\section{M-models for multivariate disease mapping} \label{sec:Mmodels}
Let us assume that the area of interest is divided into $I$ contiguous small areas and data are available for $J$ diseases. Let $O_ {ij}$ and  $E_ {ij}$ denote the number of observed and expected cases respectively in the $i$-th small area ($i=1, \ldots, I$) and for the $j$-th disease ($j=1, \ldots, J$). Conditional on the relative risks $R_{ij}$, the number of observed cases in the $i$-th area and the $j$-th disease is assumed to follow a Poisson distribution with mean $\mu_{ij}=E_{ij} \cdot R_{ij}$, that is,
\begin{eqnarray*}
	O_{ij}| R_{ij} &\sim& Poisson(\mu_{ij}=E_{ij} \cdot R_{ij}), \\
	\log \mu_{ij}&=&\log E_{ij}+\log R_{ij}.
\end{eqnarray*}
Here $E_{ij}$ is computed using indirect standardization as $E_{ij}=\sum_{k}n_{ijk}\cdot m_{jk}$, where $k$ is the age-group,
$n_{ijk}$ is the population at risk in area $i$ and age-group $k$ for the $j$-th disease, and $m_{jk}$ is the overall mortality (or incidence) rate of the $j$-th disease in the total area of study for the $k$-th age group.
The log-risk is modelled as
\begin{equation} \label{log.risk}
	\log R_{ij}=\alpha_j + \theta_{ij},
\end{equation}
where $\alpha_j$ is a disease-specific intercept and $\theta_{ij}$ is the spatial effect of the $i$-th area for the $j$-th disease.
Following the work by \cite{botella2015unifying}, we rearrange the spatial effects into the matrix $\Ttheta=\lbrace \theta_{ij}: i=1, \ldots, I; j=1, \ldots, J \rbrace$ to better comprehend the dependence structure.
The main advantage of the multivariate modelling is that dependence between the spatial patterns of the different diseases can be included in the model, so that latent associations between diseases can help to discover potential risk factors related to the phenomena under study. These unknown connections can be crucial to a better understanding of complex diseases such as cancer.

The potential association between the spatial patterns of the different diseases are included in the model considering the decomposition of $\Ttheta$ as
\begin{equation} \label{M.model}
	\Ttheta = \Phi \mathbf{M},
\end{equation}
where $\mathbf{\Phi}$ and $\mathbf{M}$ deal with dependency within and between diseases respectively. We refer to Equation~\eqref{M.model} as the M-model. In the following, we briefly describe the two components of the M-model.

The matrix $\mathbf{\Phi}$ is a matrix of order $I \times K$ and it is composed of stochastically independent columns that are distributed following a spatially correlated distribution.
Usually, as many spatial distributions as diseases are considered, that is, $K=J$, although $J$ and $K$ may be different \citep[see][for a discussion]{corpas2019convenience}.
To deal with spatial dependence, different spatial priors have been considered in the literature, most of them based on the well known intrinsic conditional autoregressive (iCAR) prior \citep{besag1974spatial}. Namely, the proper CAR (pCAR), a proper version of the iCAR; the \cite{besag1991bayesian} prior (BYM), which combines iCAR  and exchangeable random effects; the \cite{leroux1999estimation} prior (LCAR) that models spatially structured and spatially unstructured variability through a weighted sum of the iCAR precision matrix and the identity, or a modified version of the BYM model denoted as BYM2 \citep{dean2001detecting,riebler2016intuitive}. In summary, the columns of $\Pphi$ follow multivariate normal distributions with mean $\mathbf{0}$ and precision matrix $\Oomega$ whose expression depends on the spatial prior. In this paper, we consider the iCAR prior for the columns of $\Pphi$, and hence the precision matrix is $\Oomega_{\mathrm{iCAR}} = \tau (\mathbf{D}_w - \mathbf{W}) = \tau \mathbf{Q}$, where $\mathbf{W}=(w_{il})$ is the spatial binary adjacency matrix defined as $w_{ii}=0$, $w_{il}=1$ if the $i$-th and the $l$-th areas are neighbours (share a common border) and 0 otherwise, $\mathbf{D}_w= \mathrm{diag}(w_{1+}, \cdots, w_{I+})$, with the diagonal elements $w_{i+}$ being the number of neighbours of the $i$-th area, and $\tau$ is the precision parameter. Note that $\mathbf{Q}$ is the usual spatial neighbourhood matrix.

On the other hand, $\mathbf{M}$ is a $K \times J$ nonsingular but arbitrary matrix and it is responsible for inducing dependence between the different columns of $\Ttheta$, i.e, for inducing correlation between the spatial patterns of the diseases. In Equation~\eqref{M.model}, the cells of $\mathbf{M}$ act as coefficients, so they can be considered as coefficients of the log-relative risks on the underlying patterns captured in $\Pphi$ and treated as fixed effects with a normal prior distribution with mean 0 and a large (and fixed) variance.
Note that, assigning $N(0,\sigma)$ priors to the cells of $\mathbf{M}$ is equivalent to assigning a Wishart prior to $\mathbf{M}'\mathbf{M}$, i.e., $ \mathbf{M}'\mathbf{M} \sim Wishart(J, \sigma^{2} \mathbf{I}_J)$.
The multivariate approach allows the estimation of the correlation between the spatial patterns of the diseases, an interesting and useful feature, as a high positive correlation would support the hypotheses of common risk factors, and hence connections between diseases. 
The covariance matrix between the spatial patterns is obtained as $\mathbf{M}'\mathbf{M}$. For further details see \cite {botella2015unifying}.

For notation purposes and to incorporate the dependencies between different diseases in the model, we introduce the $\mathrm{vec}(\cdot)$ operator. Let $\mathbf{A}=(\mathbf{A}_1,\ldots,\mathbf{A}_J)$ be an $I \times J$ matrix with $I\times 1$ columns $\mathbf{A}_j$, for $j=1,\ldots,J$. The $\mathrm{vec}(\cdot)$ operator transforms $\mathbf{A}$ into an $IJ\times 1$ vector by stacking the columns one under the other, that is, $\mathrm{vec}( \mathbf{A} )=(\mathbf{A}'_1,\ldots,\mathbf{A}'_J)'$.
Using this notation, the multivariate Model~\eqref{log.risk} can be expressed in matrix form as
\begin{equation} \label{mat.log.risk}
	\log \mathbf{R} = \left( \mathbf{I}_J \otimes \mathbf{1}_I \right) \aalpha +  \mathrm{vec} \left( \Ttheta \right),
\end{equation}
where $\aalpha = (\alpha_1,\ldots,\alpha_J)'$, $\mathbf{R}=(\mathbf{R}'_1,\ldots,\mathbf{R}_J)'$, $\mathbf{R}_j = (R_{1j},\ldots,R_{Ij})'$, $j=1,\ldots,J$, and $\mathbf{I}_J$ and $\mathbf{1}_I$ are the $J \times J$ identity matrix and a column vector of ones of length $I$ respectively.
Then, once the between-diseases dependencies are incorporated into the model, the resulting prior distributions for $\mathrm{vec} \left( \Ttheta \right)$ with Gaussian kernel has a precision matrix given by
\begin{equation} \label{prior.vec.theta}
\Oomega_{\mathrm{vec}(\Ttheta)} =
	\left( \mathbf{M}^{-1} \otimes \mathbf{I}_I \right) \:
	\mathrm{Blockdiag}(\Oomega_{1},\ldots,\Oomega_{J}) \:
	\left( \mathbf{M}^{-1} \otimes \mathbf{I}_I \right)'.
\end{equation}
Recall that this precision matrix accounts for both within and between-disease dependencies: the $\Oomega_{1},\ldots,\Oomega_{J}$ matrices control the within-diseases spatial variability and the matrix $\mathbf{M}$ captures the between-diseases variability.
Note that if $\Oomega_{1} = \ldots = \Oomega_{J}= \Oomega_{w}$, the covariance structure is separable and can be expressed as $\Oomega_{\mathrm{vec}(\Ttheta)}^{-1}=\Oomega_{b}^{-1} \otimes \Oomega_{w}^{-1}$, where $\Oomega_{b}^{-1}=\mathbf{M}'\mathbf{M}$ and $\Oomega_{w}^{-1}$ are the between- and within-disease covariance matrices, respectively. Note that in our case $\Oomega_{w}^{-1}=\Oomega_{\mathrm{iCAR}}^{-1}$.
This M-model based framework includes both separable and non-separable covariance structures, and can accommodate different spatial dependency structures with different within-disease covariance matrices.

\subsection{Model fitting, identifiability issues and prior distributions} \label{sec:Model_fitting}
Traditionally, MCMC techniques have been used for Bayesian model fitting and inference. However,
despite the advances in research, it is widely acknowledged that MCMC techniques can be computationally very demanding.  The INLA approach \citep[see][]{rue2009approximate} has turned out to be very popular in recent years. It is designed for latent Gaussian fields and is based on integrated nested Laplace approximations and numerical integration.
Many models used in practice are implemented in \texttt{R-INLA} \citep{LindRue2015}, and others can be implemented by means of generic functions with some extra-programming work. The M-model based approach is not directly available in R-INLA, but it can be implemented using the \textsf{'rgeneric'} construct \citep[see for example][]{vicente2020SERRA}. In this paper, we use INLA for model fitting and inference.

Spatial models usually present identifiability issues which are generally overcome using sum-to-zero constraints on the spatial random effects \citep[see][for details]{eberly2000identifiability, goicoa2018spatio}. In the multivariate setting, these constraints are considered for all the diseases in the model. Additionally, the M-models bring about new identifiability issues. As pointed out by \cite{botella2015unifying}, any orthogonal transformation of the columns of $\Pphi$ and of the rows of $\mathbf{M}$ in Equation \eqref{M.model}, causes an alternative decomposition of $\Ttheta$, and therefore neither  $\Pphi$ nor $\mathbf{M}$ are identifiable and inference on them should be precluded. However, $\Ttheta$ and the covariance matrix $\mathbf{M}'\mathbf{M}$ are perfectly identifiable, so inference is confined to those quantities.
It is worth noting that the decomposition of the between-diseases covariance matrix as $\Oomega_{b}^{-1}=\mathbf{M}'\mathbf{M}$ avoids dependence on the order in which the diseases are introduced into the model, but it leads to an overparameterization problem. In the M-model proposal, $J \times J$ parameters are used to estimate the covariance matrix even though only $J\times (J+1)/2$ parameters are required. In their paper, \cite{botella2015unifying} put independent normal priors $N(0,\sigma^2)$ on each entry of the matrix $\mathbf{M}$ and they show that this is equivalent to assigning a Wishart prior to the covariance matrix, i.e., $\mathbf{M}'\mathbf{M} \sim Wishart(J, \sigma^{2} \mathbf{I}_J)$.
To avoid the overparameterization of the covariance matrix we propose to use the Barlett decomposition \citep[see, for example,][]{pena2022}. In more detail, if $\Oomega_{b}^{-1}$ is the $J \times J$  between-disease covariance matrix with $\Oomega_{b}^{-1}\sim \mathrm{Wishart}(\upsilon, \mathbf{V})$, then the Bartlett decomposition of $\Oomega_{b}^{-1}$ is the factorization
\begin{equation*}
	\Oomega_{b}^{-1}=\mathbf{L}\mathbf{A}\mathbf{A}^{'}\mathbf{L}^{'}
\end{equation*}
where $\mathbf{L}$ is the Cholesky factor of $\mathbf{V}$, and
\begin{eqnarray}
\label{Cholesky_factor}
	\mathbf{A}=
		\begin{bmatrix}
			c_1    & 0      & 0   & \cdots & 0\\
			n_{21} & c_2    & 0   & \cdots & 0\\
			n_{31} & n_{32} & c_3 & \cdots & 0\\
			\vdots & \vdots & \vdots & \ddots & \vdots\\
			n_{J1} & n_{J2} & n_{J3} & \cdots & c_J\\
		\end{bmatrix},
\end{eqnarray}
whose diagonal elements are independently distributed as $\chi^2$ random variables and the off-diagonal elements are independently distributed as normal random variables. More precisely, $c^2_j \sim \chi^2_{\upsilon-j+1}$ and $n_{jl} \sim \mathrm{N}(0,1)$ for $j,l=1,\ldots,J$ with $j>l$.
Using this decomposition, only $J\times (J+1)/2$ hyperparameters (cells of $\mathbf{A}$) are needed to estimate the covariance matrix $\Oomega_{b}^{-1}$. Note that if $\mathbf{V}=\mathbf{I}_J$, then $\mathbf{L}=\mathbf{I}_J$. Finally, to avoid order dependence with the diseases, we introduced $\mathbf{M}$ into Equation~\eqref{prior.vec.theta} as the eigen-decomposition of $\Oomega_{b}^{-1}$.
\cite{Chung2015} consider a family of Wishart densities for the prior of the covariance matrix and recommend the use of $\upsilon=J+2$ degrees of freedom to make the prior a little bit more informative. In this work we follow this recommendation. Details on how to implement this in R-INLA can be found in Appendix~\ref{sec:AppendixA}.


\section{Scalable Bayesian models for high-dimensional multivariate diseases mapping} \label{sec:high_dim}
The M-model approach can be computationally intensive or even unfeasible when the number of areas ($I$) is very large.
This limitation highlights the need for new methods. 
Here, we propose to use a divide and conquer strategy partitioning the spatial domain ($\mathfrak{D}$) into $D$ subregions, so that local multivariate spatial models can be simultaneously fitted in the different subregions. In each subregion, we consider the prior distribution with Gaussian kernel and precision matrix given in Equation~\eqref{prior.vec.theta} to deal with within-disease spatial variation and between-disease correlations.
%

\subsection{Disjoint models} \label{sec:disjoint}
A natural way to think of partitions is to consider subregions based on administrative subdivisions of the area of interest, for example provinces, states or counties. Once we have a partition of the spatial domain $\mathfrak{D}$, each geographic unit must belong to a single subregion, i.e. $\mathfrak{D}=\cup_{d=1}^{D}\mathfrak{D}_{d}$ where $\mathfrak{D}_{i} \cap \mathfrak{D}_{j} = \varnothing$ for $i\neq j$.
Then, the log-risks of the models in each subregion $d$ ($d=1,\ldots,D$) are expressed in matrix form as
\begin{eqnarray}\label{part.log.risk}
\log \mathbf{R}^{(d)} &=&
	\left( \mathbf{I}_J \otimes \mathbf{1}_{I_{d}} \right) \aalpha^{(d)} + \mathrm{vec} \left( \Ttheta^{(d)} \right), \\ \nonumber
\mathrm{vec} \left( \Ttheta^{(d)} \right) &\sim &
	\mathrm{N}\left(\mathbf{0}, \Oomega_{\mathrm{vec}\left( \Ttheta^{(d)} \right) } \right) \\
\Oomega_{\mathrm{vec}\left( \Ttheta^{(d)} \right)} &=& \nonumber
	\left[ \left( \mathbf{M}^{(d)} \right)^{-1} \times \mathbf{I}_{I_{d}} \right] \:
	\mathrm{Blockdiag}\left(\Oomega^{(d)}_{1},\ldots,\Oomega^{(d)}_{J} \right) \:
	\left[ \left( \mathbf{M}^{(d)} \right)^{-1} \times \mathbf{I}_{I_{d}} \right]'
\end{eqnarray}
where for each subregion $d$, $\aalpha^{(d)}= (\alpha^{(d)}_1,\ldots,\alpha^{(d)}_J)'$ and $\alpha^{(d)}_j$ is a disease-specific intercept, $\mathbf{I}$ is the identity matrix,  $\mathbf{R}^{(d)}=\left( \mathbf{R}^{(d)'}_{1}, \cdots, \mathbf{R}^{(d)'}_{J} \right)'$, and each $\mathbf{R}^{(d)}_{j}=(R_{1j}^{(d)},\ldots, R_{Ij}^{(d)})'$ is the vector of relative risks corresponding to disease $j$ within the subregion $d$. Finally, $\mathbf{1}_{I_{d}}$ is a column vector of ones of length $I_{d}$ (the number of areas within partition $d$), $I=\sum_{d=1}^{D}I_d$, and $\Ttheta^{(d)}=\lbrace \theta^{(d)}_{ij}: i=1, \ldots, I_d; j=1, \ldots, J \rbrace$ is the matrix of spatial effects in partition $d$ including both within and between-disease dependence structure.
%
In more detail, this model can be expressed as
\[
\begin{pmatrix}
	\log \mathbf{R}^{(1)} \\
	\vdots \\
	\log \mathbf{R}^{(d)} \\
	\vdots \\
	\log \mathbf{R}^{(D)} \\
\end{pmatrix}
	=
\mathbf{I}_J \otimes	
\begin{pmatrix}
	\mathbf{1}_{I_{1}} \\
		& \ddots	\\
		& & \mathbf{1}_{I_{d}} \\
		& & &\ddots	\\
		& & & & \mathbf{1}_{I_{D}} \\
\end{pmatrix}
\begin{pmatrix}
	\aalpha^{(1)} \\
	\vdots \\
	\aalpha^{(d)} \\
	\vdots \\
	\aalpha^{(D)} \\
\end{pmatrix}
	+
\begin{pmatrix}
	\mathrm{vec} \left( \Ttheta^{(1)} \right) \\
	\vdots \\
	\mathrm{vec} \left( \Ttheta^{(d)} \right) \\
	\vdots \\
	\mathrm{vec} \left( \Ttheta^{(D)} \right) \\
\end{pmatrix}	
\]
where the precision matrix of the multivariate normal random effect vector \sloppy $\left( \mathrm{vec} \Ttheta^{(1)'}, \ldots, \mathrm{vec} \Ttheta^{(D)'} \right)'$ is
a block-diagonal matrix of dimension $IJ \times IJ$ whose blocks correspond to the precision matrices $\Oomega_{\mathrm{vec}\left( \Ttheta^{(d)} \right)}$, $d=1,\ldots,D$. 
Having considered a partition of the spatial domain $\mathfrak{D}$, the full domain log-risk is just the union of the posterior estimates of each subregion, i.e., $\log \mathbf{R} =\left( \log \mathbf{R}^{(1)'},\cdots,\log \mathbf{R}^{(D)'} \right)'$.

\subsection{Models with overlapping partitions} \label{sec:k_order}
Disjoint partitions, such as the one considered in the previous subsection, might suffer from border effects as areas in the boundary of a given partition would not borrow information from neigbouring areas from a contiguous subdivision. Consequently, the risk estimates in those areas may not be correct.
This inconvenience can be solved by considering an alternative modelling approach in which $k$-order neighbours are added to each subregion of the partition, so that border areas have neighbours from other subregion of the partition.
In this case, the entire spatial region $\mathfrak{D}$ is divided into a set of overlapping subregions and some small areas will belong to more than one of such subdivisions, i.e., $\mathfrak{D}=\cup_{d=1}^{D}\mathfrak{D}_{d}$ and $\mathfrak{D}_{i} \cap \mathfrak{D}_{j} \neq \varnothing$ for neighbouring subregions.
Similar to the disjoint Model (\ref{part.log.risk}), $D$ submodels will be simultaneously fitted. However, as $\sum_{d=1}^{D} I_{d}>I$, the final risk $\mathbf{R}=(\mathbf{R}'_1,\ldots,\mathbf{R}'_J)'$ with $\mathbf{R}'_j = (R_{1j},\ldots,R_{Ij})'$, $j=1,\ldots,J$, is no longer the union of the posterior estimates obtained for each submodel as areas located in the borders of the spatial partition would have more than one estimated posterior distribution.

Two different strategies can be considered to obtain a unique posterior estimate of the relative risk for those areas in more than one subregion. \cite{orozco2021scalable} propose to calculate the mixture distribution of the estimated posterior probability density functions of the relative risks in the different subdivisions, with weights proportional to the conditional predictive ordinate (CPO) values \citep{pettit1990conditional}.
To compute the mixture, suppose that area $i$ belongs to $m(i)$ subregions of the spatial domain $\mathfrak{D}$ and let $f^{(1)}_{ij}(x),\cdots,f^{(m(i))}_{ij}(x)$ be the posterior estimates of the probability density functions of the $j$-th disease in the $i$-th area. Then the mixture distribution of $R_{ij}$ can be written as
\begin{equation*}
f_{ij}(x)=\sum_{k=1}^{m(i)} w_{k} f^{(k)}_{ij}(x), \quad \mbox{with} \quad w_k = \dfrac{CPO_{ij}^k}{\sideset{}{_{k=1}^{m(i)}}\sum CPO_{ij}^k}
\end{equation*}
where $CPO_{ij}^k$ is the conditional predictive ordinate of area $i$ and disease $j$ obtained in partition $k$, so that $w_{k} \geq 0$ and $\sum_{k=1}^{m(i)} w_{k}=1$ \citep[see for example][]{lindsay1995mixture, fruhwirth2006finite}.

More recently, \cite{orozco2022} consider using the posterior marginal distribution of the relative risk estimated from its original partition. Based on the results obtained from a simulation study, they show that this strategy outperforms the use of mixture distributions in terms of risk estimation accuracy and true positive/negative rates. In this paper, this is also the default strategy used to obtain unique posterior distributions for each relative risk $R_{ij}$.

\subsection{Between-disease correlations and variance parameters} \label{sec:corre_fix}
In addition to enlarge the effective sample size and improving smoothing by borrowing information from different diseases, one of the main advantages of multivariate disease mapping models is that they take into account correlations between the  spatial patterns of the different diseases, that is, they reveal connections between diseases. Fitting a single multivariate model to the region of interest provides correlations between the diseases in the whole study domain thus revealing overall relationships. In addition, it also provides the diagonal elements of the between-disease covariance matrix, hereafter referred to as variance parameters. In the case of separable covariance structures (the kronecker product of between and within disease covariance matrices) these parameters control the amount of smoothing within diseases.
%
By dividing the spatial domain into subregions, we obtain the posterior distributions of these parameters in each of the subdivisions.  In addition, we are able to retrieve the between disease correlations and variances for the whole region. Hence, partition models provide extra information as they give insight about local connections between the diseases in the subdivisions (which in general are administrative divisions) and the global connection in the whole study region.

To obtain global estimates of the parameters of interest in the overall study domain from the partition models, we adapt the consensus Monte Carlo (CMC) algorithm originally proposed by \cite{scott2016consensus}. The idea behind consensus Monte Carlo is to divide the data into shards (in our case, the shards corresponds to different subdivisions of the spatial domain), give each shard to a worker machine which does a full Monte Carlo simulation from a posterior distribution given its own data, and then combine the posterior simulations from each worker (or submodel) to produce a set of global draws representing the consensus belief among all the workers. Here, we briefly describe how to adapt the ideas behind the CMC algorithm to our case.

Let ${\ppsi}=({\rrho},{\ssigma}^2)^{'}$ denotes the vector with the parameters of interest where ${\rrho}=(\rho_{12},\ldots,\rho_{J-1,J})^{'}$ contains the between-disease correlations and ${\ssigma}^2=(\sigma^2_1,\ldots,\sigma^2_J)^{'}$ are the diagonal elements of the between-disease covariance matrix, and let ${\psi}_{kd}$ denote the local estimate of the $k$-th parameter of ${\ppsi}$ in each subdomain $\mathfrak{D}_{d}$, $d=1,\ldots,D$.
We first extract samples of size $S$ from the posterior marginal estimates of $\psi_{kd}$
 denoted as $\psi_{kd}^s$ for $k=1,\ldots,J \times (J+1)/2$, $d=1,\ldots,D$ and $s=1,\ldots,S$. Then, we combine the draws using weighted averages
\begin{equation*}
\tilde{\psi}_k^s=\sum\limits_{d=1}^D w_d \psi_{kd}^s, \quad \mbox{for } s=1,\ldots,S
\end{equation*}
where $w_d$ are normalized weights inversely proportional to the posterior marginal variances of $\psi_{kd}$. Finally, we approximate the posterior marginal density function of the parameter $\psi_k$ from the combined draws $\tilde{\psi}_k^s$.

\subsection{Model selection criteria} \label{sec:criteria}
Two of the most widely used criteria to compare Bayesian models are the deviance information criterion (DIC) \citep{spiegelhalter2002bayesian} and the Watanabe-Akaike information criterion (WAIC) \citep{watanabe2010asymptotic}.
However, with partition models, it is not straightforward to get these quantities as we fit as many models as subdivisions. Hence, a procedure to estimate these quantities from the scalable models described in Sections \ref{sec:disjoint} and \ref{sec:k_order} is required.

Extending the ideas in \cite{orozco2021scalable} to the multivariate framework, we compute approximate DIC values by drawing samples from the posterior marginal distribution of the Poisson means.
Denoting by $\mathbf{C}^{s}$, $s=1,...,S$, to the posterior simulations of $\mu_{ij}=E_{ij}\cdot R_{ij}$ (the mean of the Poisson distribution), approximate values of the mean deviance ($\overline{D(\mathbf{C})}$) and the deviance of the mean ($D(\overline{\mathbf{C}})$) can be respectively calculated as
\begin{eqnarray*}
\overline{D(\mathbf{C})} =
	\dfrac{1}{S}\sum_{s=1}^{S} - \log \left( p(\mathbf{O} \vert \mathbf{C}^{s}) \right) ; \quad
D(\overline{\mathbf{C}}) =
	-2\log \left( p(\mathbf{O} \vert \overline{\mathbf{C}}) \right),
	\: \mathrm{with} \:
	\overline{\mathbf{C}} = \dfrac{1}{S}\sum_{s=1}^{S} \mathbf{C}^{s}.
\end{eqnarray*}
where $p(\mathbf{O}|\cdot)$ denotes the likelihood function of a Poisson distribution. Then, the DIC is obtained as

\[
\mathrm{DIC} =
	2 \, \overline{D(\mathbf{C})} - D(\overline{\mathbf{C}}).
\]

Similary, approximate WAIC values are computed as \citep[see][]{gelman2014understanding}
\begin{equation*}
\mathrm{WAIC} = -2 \sum_{i=1}^{I}\sum_{j=1}^{J} \log \left(
		\dfrac{1}{S}\sum_{s=1}^{S} p(O_{ij} \vert\mathbf{C}^{s}) \right) +
	2 \sum_{i=1}^{I}\sum_{j=1}^{J} \mathrm{var} \left[ \log \left( p(O_{ij} \vert \mathbf{C}^{s}) \right) \right].
\end{equation*}

\section{Simulation study} \label{sec:simulation}
We conduct a simulation study to compare the performance of the different M-models described in Section~\ref{sec:Mmodels}. Specifically, our interest relies on comparing the fit of a single model to the whole domain (hereafter referred to as the global model) and the partition models,  in terms of parameter estimates and relative risk estimation accuracy. The $I=7907$ municipalities of continental Spain and $J=3$ diseases are used as the simulation template, as this imitates the case study presented in Section~\ref{sec:Case_study}.

Two different scenarios have been considered to recover the possible underlying generating process of spatially correlated disease risks. In the first scenario, samples are generated from a fixed covariance structure based on the spatial neighbourhood graph of the whole area under study, that is, the global model is used as the generating model. In contrast, in the second scenario, independent samples for each partition (Spanish Autonomous Regions, see \autoref{fig:Map_CCAA} in  Appendix~\ref{sec:AppendixB}) are generated using the covariance structures of the partition, that is, the Disjoint model is used as the data generating mechanism. Further details are given below.


\subsection{Data generation} \label{sec:simulation_data}
As one of the main advantages of the joint modelling of several responses is to analyze the relationships between different diseases in terms of correlations between spatial patterns, we are interested in evaluating how well these parameters are estimated when using the multivariate spatial models described in this paper.
To do this, we start by sampling from a multivariate normal distribution with precision matrix equal to $\Oomega_{\mathrm{vec}(\Ttheta)}=\Oomega_b \otimes \Oomega_{iCAR}$, by fixing the elements of the between-disease covariance matrix as
\begin{equation*}
\Omega^{-1}_b =
\begin{pmatrix} \sigma_1 & & \\ & \sigma_2 & \\ & & \sigma_3 \end{pmatrix}
\begin{pmatrix}
    1 & \rho_{12} & \rho_{13} \\
    \rho_{21} & 1 & \rho_{23} \\
    \rho_{31} & \rho_{32} & 1 \\
\end{pmatrix}
\begin{pmatrix} \sigma_1 & & \\ & \sigma_2 & \\ & & \sigma_3 \end{pmatrix}
=
\begin{pmatrix}
    \sigma_1^2 & \sigma_{12} & \sigma_{13} \\
    \sigma_{21} & \sigma_2^2 & \sigma_{23} \\
    \sigma_{31} & \sigma_{32} & \sigma_3^2 \\
\end{pmatrix}
\end{equation*}
where $\sigma^2_j$ are variance parameters, and $\rho_{kl}=\rho_{lk}$ are between-disease correlation coefficients. Note that $\sigma_{kl}$ denotes the covariances between each pair of diseases. Then, for each sample of $\Ttheta^{r}$, $r=1,\ldots,100$ we compute the relative risks $R_{ij}^r$ following Equation~\eqref{mat.log.risk}. Finally, we generate $O_{ij}$ counts for area $i$ and disease $j$ using a Poisson distribution with mean $\mu_{ij}^r = E_{ij} \cdot R_{ij}^r$, where $E_{ij}$ are the expected number of cases of our case study data (lung, colorectal and stomach cancer mortality in Spanish males).


In Scenario 1, the neighbourhood graph of all the 7907 municipalities is used to define the spatial precision matrix $\Oomega_{iCAR}$. In addition, we fix the parameters of the between-disease covariance matrix as $\sigma_1^2=0.25$, $\sigma_2^2=0.16$, $\sigma_3^2=0.09$, $\rho_{12}=0.7$, $\rho_{13}=0.5$ and $\rho_{23}=0.1$.
In Scenario 2, $D=15$ independent samples are generated from multivariate Normal distributions with precision matrices equal to $\Oomega_{\mathrm{vec}(\Ttheta^{d})}= \Oomega_b^{(d)} \otimes \Oomega_{iCAR}^{(d)}$, where $\Oomega_{iCAR}^{(d)}$ is the spatial precision matrix of the areas within subdomain $d=1,\ldots,D$, and different between-disease covariance matrices $\Oomega_b^{(d)}$ are considered en each subdivision. Here, the variance parameters are fixed to $\sigma^2_1=0.5$, $\sigma^2_2=0.4$ and $\sigma^2_3=0.3$, while similar values to the ones estimated with the partition models in the case study presented in the next section are used as correlation coefficients (see \autoref{tab:Scenario2_k0} in Appendix~\ref{sec:AppendixB}). Note that we increase the variance parameters in Scenario 2 to get stronger smoothing effects in each subdivision.

\subsection{Results: Scenario 1} \label{sec:scenario1_results}
\autoref{tab:scenario1_parameters} compares the true values of model parameters in Scenario 1 (variance parameters and correlation coefficients) against average values of posterior mean estimates over the 100 simulated data sets. In addition, estimated standard errors, simulated standard errors (derived from the sample variance of the parameter estimates) and empirical coverages of the 95\% credible intervals are also displayed. Note that for the partition models, these posterior marginal distributions are obtained by using the CMC algorithm described in Section~\ref{sec:corre_fix}. In term of model parameters, multivariatie models give very accurate estimates of the real values, both in terms of posterior mean and posterior standard deviation estimates (note that nearly identical values are obtained from estimated and simulated standard errors). As expected, slightly better results are obtained when fitting the global model, as this is the true generating model in Scenario 1. Regarding partition models, the higher the neighbourhood order, the more similar the CMC estimates of the correlation coefficients are to those of the global model.

\autoref{tab:scenario1_DIC} displays average values of model selection criteria (posterior mean deviance $\overline{D(\ttheta)}$, effective number of parameters $p_D$, DIC and WAIC) for the global and the scalable models, as well as the accuracy of the relative risk estimates quantified by the mean absolute relative bias (MARB), the mean relative root mean square errors (MRRMSE) and empirical coverages of the 95\% credible intervals for the risks.
Note that the MARB and MMRMSE are defined for each small area $i$ and disease $j$ as
\begin{equation*}
\mbox{MARB}_{ij} = \left| \frac{1}{100}\sum_{r=1}^{100} \dfrac{\hat{R}_{ij}^{(r)}-R_{ij}^{(r)}}{R_{ij}^{(r)}} \right|
\quad \mbox{and} \quad
\mbox{MMRMSE}_{ij} = \sqrt{\frac{1}{100} \sum_{r=1}^{100} \left(\dfrac{\hat{R}_{ij}^{(r)}-R_{ij}^{(r)}}{R_{ij}^{(r)}} \right)^2}
\end{equation*}
where $R_{ij}^{(r)}$ and $\hat{R}_{ij}^{(r)}$ denote the true value and the posterior median estimate of the relative risks for the $s$-th data set ($s=1,\ldots,100$). Model selection criteria point towards partition models, though differences are mild. Regarding MARB, MMRMSE and 95\% coverage values, differences between the global and the partition models are practically negligible.

\begin{table}[!ht]
\caption{Average values of posterior mean, posterior standard deviation (SD), simulated standard errors (sim) and empirical coverage of the 95\% credible intervals (EC) for model parameters based on 100 simulated data sets for Scenario 1.}
\label{tab:scenario1_parameters}
\begin{center}
\renewcommand{\arraystretch}{1.15}
\begin{tabular}{cc|rrrr|rrrr}
\toprule
& & \multicolumn{4}{c|}{Global model} & \multicolumn{4}{c}{Disjoint model} \\
& & & & & & & & & \\
& True value & Mean & SD & Sim & EC & Mean & SD & Sim & EC  \\
\hline
$\sigma^2_1$ &  0.25 &  0.250 & 0.011 & 0.011 & 0.95 &  0.240 & 0.012 & 0.012 & 0.83 \\
$\sigma^2_2$ &  0.16 &  0.160 & 0.010 & 0.010 & 0.95 &  0.158 & 0.011 & 0.011 & 0.96 \\
$\sigma^2_3$ &  0.09 &  0.092 & 0.009 & 0.010 & 0.92 &  0.101 & 0.010 & 0.011 & 0.74 \\[1.ex]
$\rho_{12}$  &  0.70 &  0.700 & 0.025 & 0.026 & 0.95 &  0.690 & 0.026 & 0.029 & 0.89 \\
$\rho_{13}$  &  0.50 &  0.487 & 0.044 & 0.046 & 0.95 &  0.452 & 0.045 & 0.048 & 0.80 \\
$\rho_{23}$  &  0.10 &  0.089 & 0.059 & 0.057 & 0.96 &  0.077 & 0.057 & 0.065 & 0.95 \\
\hline\\[-2.ex]
& & \multicolumn{4}{c|}{1st-order nb model} & \multicolumn{4}{c}{2nd-order nb model} \\
& & & & & & & & & \\
& True value & Mean & SD & Sim & EC & Mean & SD & Sim & EC  \\
\hline
$\sigma^2_1$ &  0.25 & 0.241 & 0.011 & 0.012 & 0.84 & 0.239 & 0.010 & 0.012 & 0.75 \\
$\sigma^2_2$ &  0.16 & 0.159 & 0.010 & 0.010 & 0.94 & 0.155 & 0.009 & 0.010 & 0.89 \\
$\sigma^2_3$ &  0.09 & 0.100 & 0.010 & 0.010 & 0.79 & 0.097 & 0.009 & 0.010 & 0.83 \\[1.ex]
$\rho_{12}$  &  0.70 & 0.691 & 0.025 & 0.029 & 0.92 & 0.695 & 0.023 & 0.032 & 0.82 \\
$\rho_{13}$  &  0.50 & 0.461 & 0.043 & 0.051 & 0.81 & 0.468 & 0.040 & 0.048 & 0.83 \\
$\rho_{23}$  &  0.10 & 0.079 & 0.055 & 0.058 & 0.91 & 0.082 & 0.053 & 0.060 & 0.95 \\
\bottomrule
\end{tabular}
\end{center}
\end{table}

\begin{table}[!ht]
\caption{Average values of model selection criteria (mean deviance, effective number of parameters, DIC and WAIC) and risk estimation accuracy (MARB, MRRMSE and empirical coverage -EC- of the 95\% credible intervals) based on 100 simulated data sets for Scenario 1.}
\label{tab:scenario1_DIC}
\begin{center}
\renewcommand{\arraystretch}{1.15}
\begin{tabular}{l|ccccccccc}
\toprule
& \multicolumn{4}{c}{Model selection criteria} && \multicolumn{3}{c}{Risk estimation accuracy} \\
\cmidrule{2-5} \cmidrule{7-9}
 & $\overline{D(\ttheta)}$ & $p_D$ & DIC & WAIC && MARB & MRRMSE & EC \\
\midrule
Global       & 78521.9 & 3046.9 & 81568.8 & 81504.7 && 0.024 & 0.191 & 0.950 \\
Disjoint     & 78299.9 & 3329.3 & 81629.1 & 81529.2 && 0.023 & 0.196 & 0.957 \\
1st-order nb & 78407.9 & 3154.7 & 81562.6 & 81499.4 && 0.024 & 0.193 & 0.953 \\
2nd-order nb & 78454.5 & 3091.5 & 81546.0 & 81496.5 && 0.024 & 0.192 & 0.950 \\
\bottomrule
\end{tabular}
\end{center}
\end{table}


\subsection{Results: Scenario 2} \label{sec:scenario2_results}

In contrast to the previous scenario, it should be noted that in Scenario 2 we cannot compare the global estimates of the model parameters against the true values of the variance parameters and between-disease correlations, since different values have been used to generate the risk surfaces in each subdomain. However, we can compare the model's performance in terms of model selection criteria and risk estimation accuracy (see \autoref{tab:scenario2_DIC}). As expected, the Disjoint model is the one showing the best performance according to these measures, as this is the true generating model assumed for Scenario 2. Although slightly worse MARB and MRRMSE values are obtained for 1st/2nd-order neighbourhood models, the partition models still outperform the Global model.

We are also interested in analyzing if the partition models are able to recover the local between-disease covariance structures of the true generating process. In \autoref{tab:Scenario2_k0} (Appendix~\ref{sec:AppendixB}) we compare these values against the average values of posterior mean estimates of local parameters in each subdivision over the 100 simulated data sets for the Disjoint model. For almost every subdivision, very accurate estimates are obtained for both variance parameters and correlation coefficients. For the latter, the median value of the empirical coverage of the 95\% credible intervals is 0.95 (with $Q_1=0.93$ and $Q_3=0.97$). As expected, these estimates get worse as the neighbourhood order of the models increases, since the estimated local correlation structures are affected by the ones of the adjacent subdivisions. Even so, the median values of the empirical coverage of the 95\% credible intervals for the between-disease correlations are 0.89 (with $Q_1=0.84$ and $Q_3=0.92$) and 0.86 (with $Q_1=0.79$ and $Q_3=0.90$) for 1st-order and 2nd-order neighbourhood models, respectively. All the results are shown in \autoref{tab:Scenario2_k1} and \autoref{tab:Scenario2_k2} in Appendix~\ref{sec:AppendixB}.

\begin{table}[!ht]
\caption{Average values of model selection criteria (mean deviance, effective number of parameters, DIC and WAIC) and risk estimation accuracy (MARB, MRRMSE and empirical coverage -EC- of the 95\% credible intervals) based on 100 simulated data sets for Scenario 2.}
\label{tab:scenario2_DIC}
\begin{center}
\renewcommand{\arraystretch}{1.15}
\begin{tabular}{l|ccccccccc}
\toprule
& \multicolumn{4}{c}{Model selection criteria} && \multicolumn{3}{c}{Risk estimation accuracy} \\
\cmidrule{2-5} \cmidrule{7-9}
 & $\overline{D(\ttheta)}$ & $p_D$ & DIC & WAIC && MARB & MRRMSE & EC \\
\midrule
Global       & 78766.9 & 5385.6 & 84152.5 & 83894.7 && 0.062 & 0.322 & 0.947 \\
Disjoint     & 78505.8 & 5132.6 & 83638.4 & 83451.3 && 0.051 & 0.299 & 0.954 \\
1st-order nb & 78420.2 & 5465.5 & 83885.7 & 83650.9 && 0.055 & 0.314 & 0.957 \\
2nd-order nb & 78457.1 & 5460.2 & 83917.3 & 83694.3 && 0.057 & 0.317 & 0.955 \\
\bottomrule
\end{tabular}
\end{center}
\end{table}

\section{Case study} \label{sec:Case_study}
In this section we jointly analyse mortality data for lung, colorectal, and stomach cancer in men in the 7907 municipalities of mainland Spain (excluding Baleares and Canary Islands and the autonomous cities of Ceuta and Melilla) during the period 2006-2015 using the new proposal.
During the ten years of the study, a total of 162,602 deaths from lung cancer (corresponding to codes C33-C34 of the International Classification of Diseases-10), 82,967 from colorectal cancer (C17-C21) and 33,170 from stomach cancer (C16) were registered for male population of mainland Spain, which correspond to global rates of 76.48, 39.02 and 15.60 deaths per 100,000 male inhabitants, respectively.

\subsection{Model fitting and model selection} \label{sec:model_fitting}
We fit the disjoint model ($k=0$) and the $k$-order neighbourhood model for $k=1,2,3$ in R-INLA using $D=15$ subdivisions of the spatial domain. These subdivisions are also of interest as they correspond to Autonomous Regions of Spain (NUTS2 level from the European nomenclature of territorial units for statistics, shown in \autoref{fig:Map_CCAA} in Appendix \ref{sec:AppendixB}).
In these partitions, the highest value of $I_{d}$ (number of municipalities) is 2245 and corresponds to the Autonomous Region of Castilla y Le{\'o}n, a rather vast territory from central to northwestern Spain with about 5\% of the total Spanish population. Although this sub-region is large, we maintain this subdivision as it represents the administrative division of Spain into Autonomous Regions. We also fit the multivariate spatial M-models over the entire spatial domain (global model), and compare the results with those obtained with the new proposal.

Previously, univariate models were also fitted to each disease using a BYM2 spatial prior. The covariance matrix of this prior cope with both spatial structured variability and unstructured variability. Results (not shown here to conserve space) show that most of the variability is spatially structured. Since the computational cost of this prior makes it difficult its use in a multivariate setting, and most of the variability is spatially structured, we fit the joint multivariate proposal given in Equation~\eqref{part.log.risk} by considering an iCAR prior for the spatial random effects.


For the partition models, we distribute the submodels over 2 machines with four processors Intel Xeon Silver 4108 and 192GB RAM on each machine (Ubuntu 20.04.4 LTS operative system), using the simplified Laplace approximation strategy in R-INLA \citep{LindRue2015} (stable version INLA\_22.05.07, R version R-4.1.2) and simultaneously running 3 models in parallel on each machine using the \texttt{bigDM} package \citep{bigDM}.

\begin{table}[ht]
\centering
\caption{Model selection criteria and computational time, in minutes, for multivariate models with iCAR spatial prior using the simplified Laplace approximation strategy if INLA.}
\label{tab:CaseStudy_DIC}
\vspace{-0.5cm}
\begin{center}
\renewcommand{\arraystretch}{1.15}
\begin{tabular}{lrrrrrrrr}
\\[-2.ex]
\toprule
& \multicolumn{4}{c}{Model selection criteria} && \multicolumn{3}{c}{Time}\\
\cmidrule{2-5} \cmidrule{7-9}
Model & $\overline{D(\ttheta)}$ & $p_D$ & DIC & WAIC && Run & Merge & Total \\
\midrule
k=0    & 76779.7 & 2471.9 & 79252.6 & 79204.8 &&  5.4 & 0.7 &  6.1 \\
k=1    & 76894.6 & 2327.4 & 79222.0 & 79187.3 &&  6.5 & 1.1 &  7.6 \\
k=2    & 76942.0 & 2289.4 & 79231.4 & 79211.9 &&  7.7 & 1.1 &  8.8 \\
k=3    & 77007.0 & 2231.8 & 79238.8 & 79220.0 &&  8.2 & 1.1 &  9.3 \\
Global & 77186.8 & 2164.2 & 79351.0 & 79283.9 && 33.2 & $-$ & 33.2 \\
\bottomrule
\end{tabular}
\end{center}
\end{table}

\autoref{tab:CaseStudy_DIC} displays the posterior mean deviance $\overline{D(\ttheta)}$, the effective number of parameters $p_D$, the DIC, and the WAIC for the global and the scalable models together with the computing time. The total time for the scalable models is obtained by adding the running time and the merging time.
%
The running time refers to the elapsed time for all the submodels fitted with R-INLA, and the merging time refers to the combination (when necessary) of the posterior distributions of the risks, the approximation of the DIC/WAIC values, and the computation of global estimates of the between-diseases correlation coefficients using the proposed CMC algorithm.
As expected, the computational cost raises as the neighbourhood order ($k$) increases, though the scalable proposal is faster than the global model for all values of $k$. The greatest reduction in time in comparison with the global model is obtained for $k=0$, being the global model about 5.5 times slower. When the neighbourhood order increases, the difference in computing time is less pronounced. The global model is about 4.3, 3.8, and 3.6 times slower than the scalable models with $k=$1, 2, and 3, respectively.
Regarding model selection criteria, scalable Bayesian models outperform the global model. The greater reduction in DIC and WAIC is obtained for the 1st-order neighbourhood model. However, increasing the neighbourhood order may improve the between-disease correlation estimates.

\subsection{Joint analysis of male mortality from three types of cancer in Spain} \label{sec:join_analysis}
In this subsection, the spatial patterns of lung, colorectal, and stomach cancer mortality risks in men are examined in the  municipalities of continental Spain using the scalable multivariate proposal presented in Section \ref{sec:high_dim}.

\begin{figure}[!ht]
\centering
  \includegraphics[width=\textwidth]{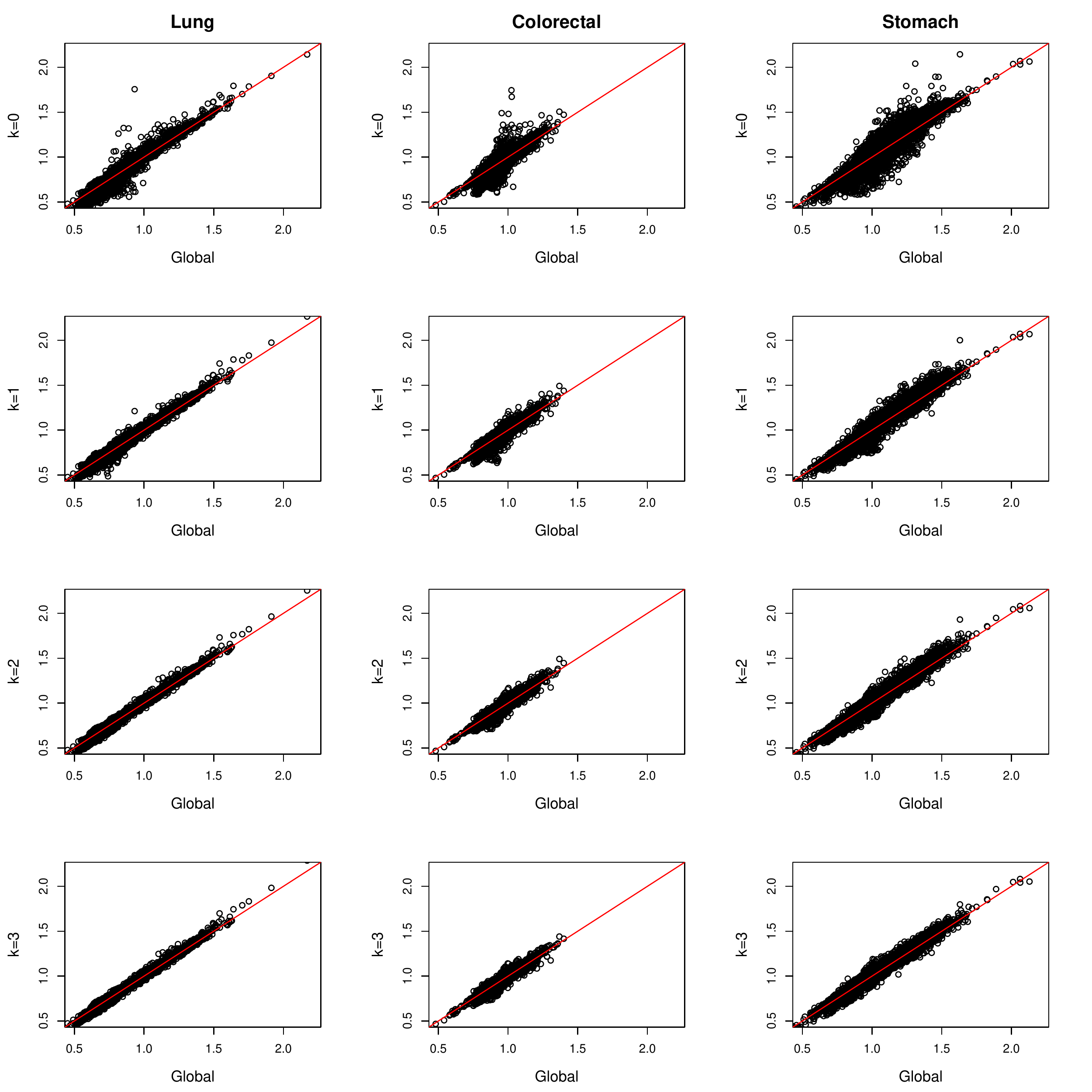}
\caption{Dispersion plots of the posterior median estimates of relative risks for lung (left column), colorectal (central column) and stomach (right column) cancer mortality data obtained with the partitioned model ($k=0,1,2,3$ from top to bottom) versus the global model.}
\label{Figure1}
\end{figure}

\begin{figure}[!ht]
\begin{center}
    \vspace{-1.5cm}
    \includegraphics[width=1.25\textwidth]{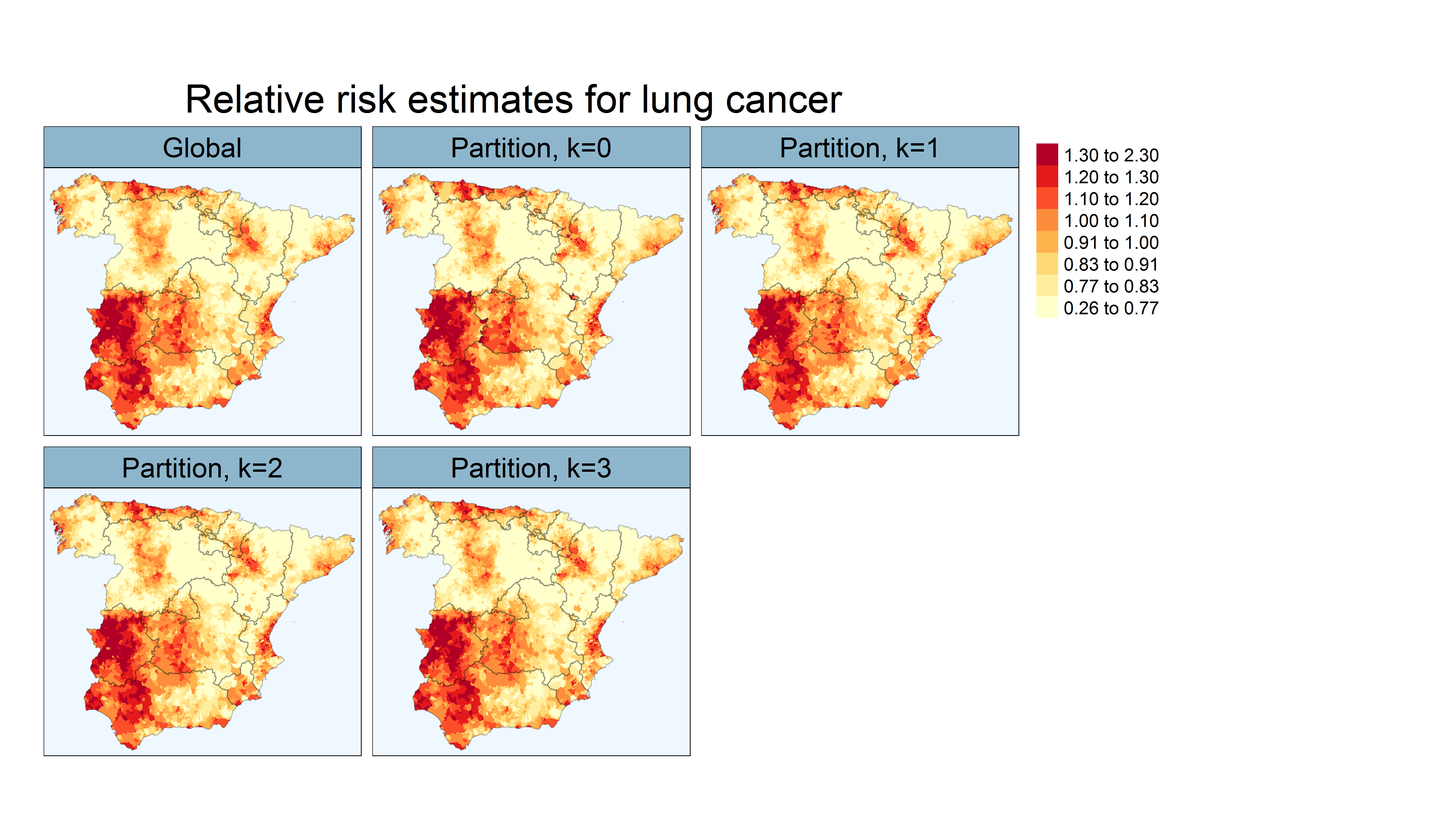}\hfill
    \vspace{-1.5cm}
    \includegraphics[width=1.25\textwidth]{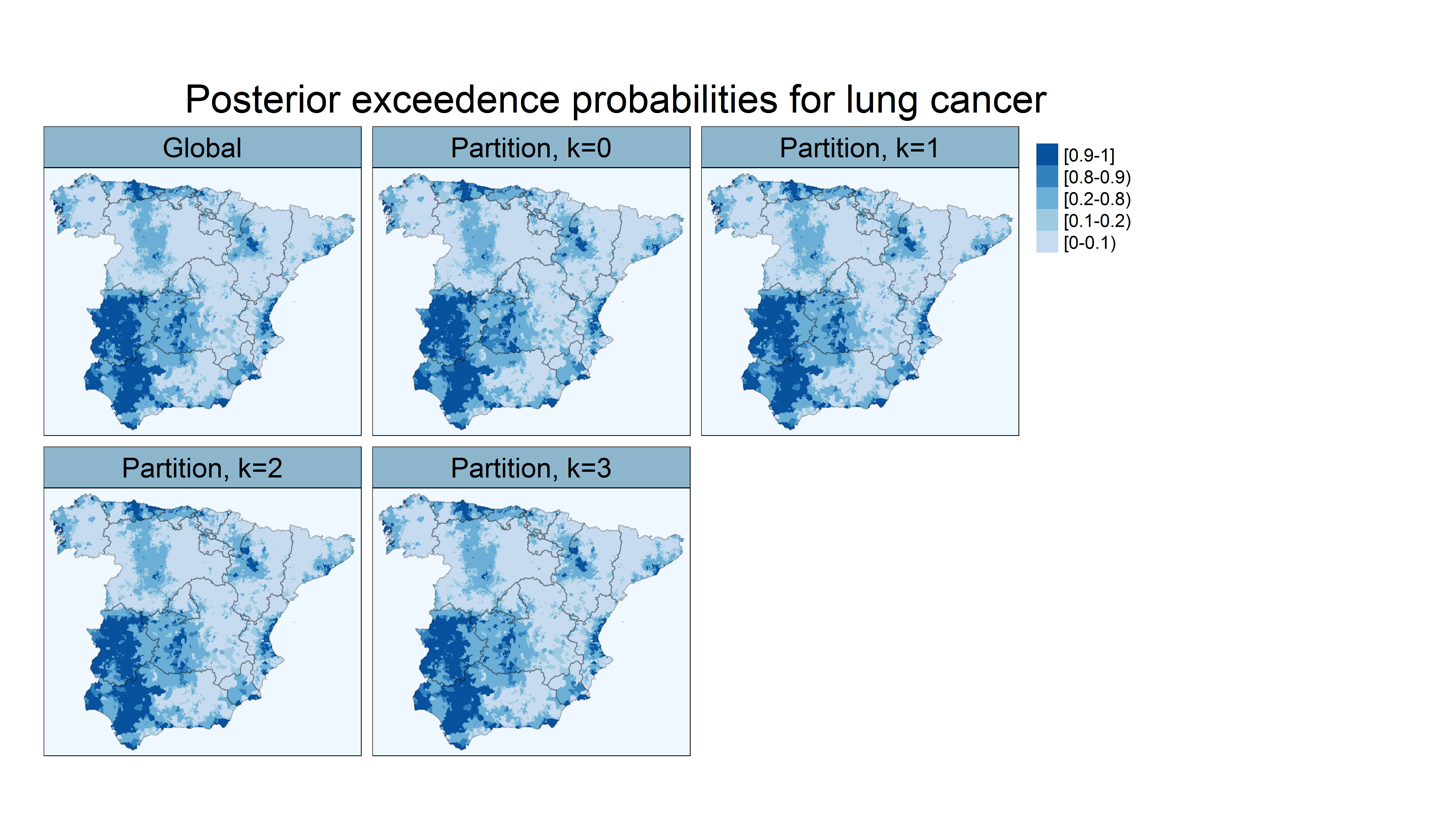}\hfill
    \vspace{-1.5cm}
\end{center}
\caption{Maps of posterior median estimates of mortality relative risk for lung cancer (top) and posterior exceedance probabilities $P(R_{ij}>1 \vert \mathbf{O})$ (bottom) in continental Spain.}
\label{Figure2}
\end{figure}

We begin with a comparison of the estimated risks obtained with the global model, the disjoint model ($k=0$) and the $k$-order neighbourhood models ($k=1$, 2 and 3). \autoref{Figure1} displays dispersion plots of the posterior median estimates of the relative risks obtained with the partitioned models versus those obtained with the global model. The left, central and right columns correspond to lung, colorectal and stomach cancer, respectively. The neighbourhood order in the partition models are represented in the different rows. The largest differences are observed between the global and the disjoint model. This is expected because areas in the border of a subdivision do not borrow strength from neighbouring areas located in a contiguous subdivision. As the neighbourhood order $k$ increases the risk estimates are more similar to the global model.
\autoref{Figure2} displays the spatial patterns of lung cancer mortality risks (top) and the posterior probabilities of risk exceedance (bottom), $P(R_{ij}>1 \vert \mathbf{O})$ , obtained with the global and the disjoint models. To save space, maps for colorectal and stomach cancer are provided in \autoref{fig:risk_icar_colorectal} and \autoref{fig:risk_icar_stomach} (Appendix~\ref{sec:AppendixB}). Though differences in risks estimates are observed in the dispersion plots, it is harder to appreciate them on the maps.

Multivariate models borrow information from nearby areas and the different diseases subject to analysis. In addition to this strength, multivariate models present additional advantages over univariate counterparts, such as the possibility of estimating correlations between the spatial patterns of the diseases.
Moderate to high correlations may suggest the existence of underlying risk factors affecting the diseases under study, which in turns implies connection between them. This information may be crucial to better understand diseases such as cancer in which known risk factors only explain a small percentage of the cases. Spatial patterns may be associated to factors like access to treatment or life style that might have an impact on mortality.

\begin{figure}[!ht]
\centering
  \includegraphics[width=\textwidth]{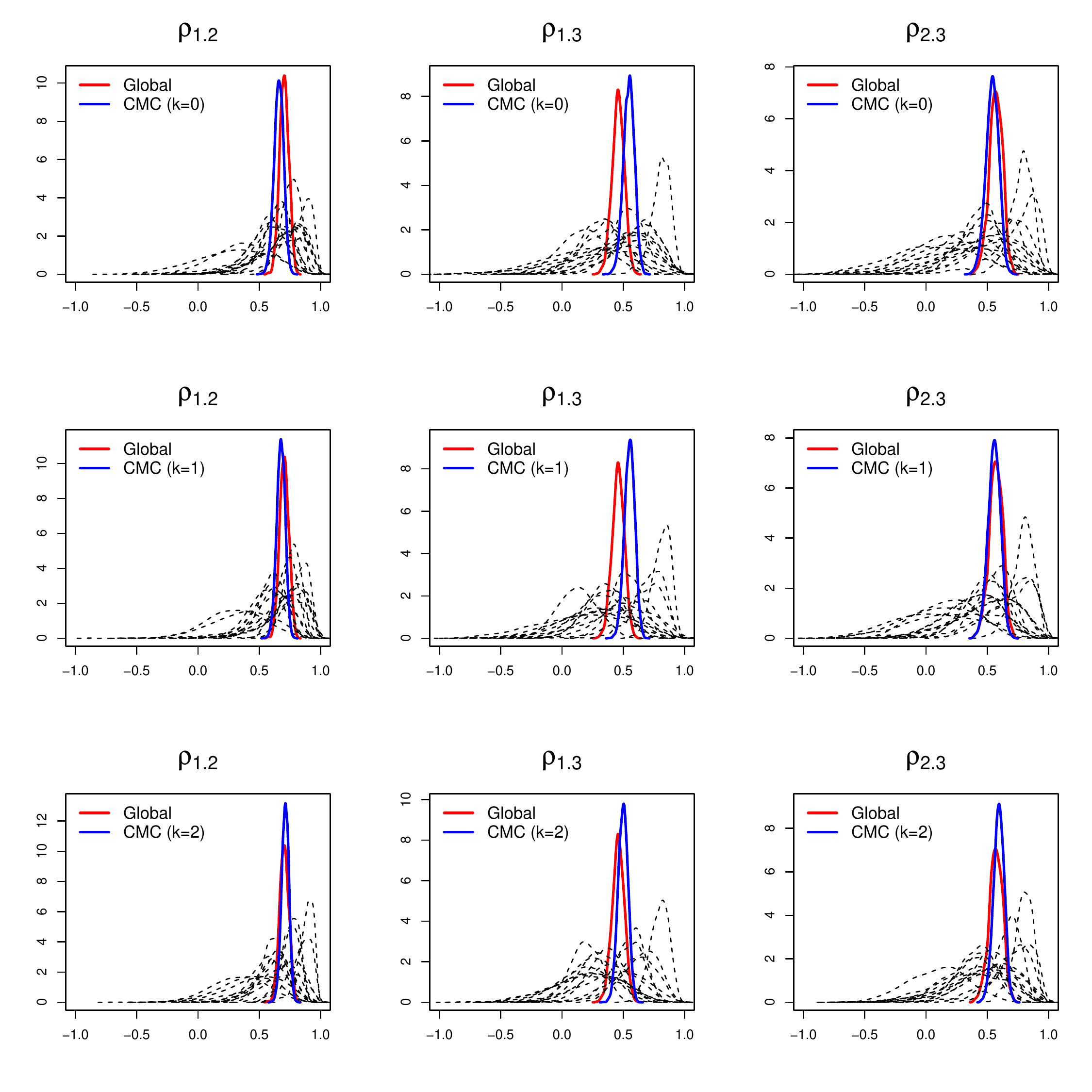}
  \vspace{-1cm}
\caption{Posterior distributions of the estimated between-disease correlations with the global, and $k=0,1,2$-order neighbourhood models, using an iCAR prior for spatial random effects.}
\label{Figure3}
\end{figure}

Posterior distributions of the between-disease correlations obtained in the different partitions with $k=0$, 1 and 2 are displayed in \autoref{Figure3} together with correlations for whole Spain obtained with the CMC algorithm and with the global model.
Here, $\rho_{1.2}$, $\rho_{1.3}$, and $\rho_{2.3}$ denote the correlation parameters between lung and colorectal, lung and stomach, and colorectal and stomach cancer, respectively.
%
Summary statistics (mean, median, mode, standard deviation, 2.5 and 97.5 percentiles) of the between-disease posterior correlations are also shown in \autoref{tab:CaseStudy_cor}. In general, the posterior distributions estimated with the CMC algorithm for the partition models are very similar to those obtained with the global model. Similar to the posterior estimates of the relative risks, closer values to the global model are observed as the neighbourhood order $k$ increases.

\begin{table}[!t]
\caption{Descriptive statistics of the estimated between-disease correlations with the global, and $k=0,1,2$-order neighbourhood models, using an iCAR prior for spatial random effects. \label{tab:CaseStudy_cor}}
\begin{center}
\renewcommand{\arraystretch}{1.15}
\begin{tabular}{llrrrrrr}
\toprule
$\rho$ & Model & mean & sd & $q_{.025}$ & $q_{.5}$ & $q_{.975}$ & mode \\
\midrule
\multirow{4}{*}{$\rho_{1.2}$}
& Global & 0.70 & 0.04 & 0.63 & 0.70 & 0.77 & 0.70 \\
& k=0    & 0.66 & 0.04 & 0.58 & 0.66 & 0.73 & 0.66 \\
& k=1    & 0.68 & 0.04 & 0.60 & 0.68 & 0.74 & 0.68 \\
& k=2    & 0.71 & 0.03 & 0.65 & 0.71 & 0.77 & 0.71 \\
\midrule
\multirow{4}{*}{$\rho_{1.3}$}
& Global & 0.46 & 0.05 & 0.36 & 0.46 & 0.55 & 0.46 \\
& k=0    & 0.55 & 0.05 & 0.46 & 0.55 & 0.63 & 0.55 \\
& k=1    & 0.55 & 0.04 & 0.47 & 0.56 & 0.63 & 0.56 \\
& k=2    & 0.50 & 0.04 & 0.42 & 0.50 & 0.57 & 0.50 \\
\midrule
\multirow{4}{*}{$\rho_{2.3}$}
& Global & 0.57 & 0.05 & 0.46 & 0.57 & 0.67 & 0.57 \\
& k=0    & 0.54 & 0.05 & 0.43 & 0.54 & 0.64 & 0.54 \\
& k=1    & 0.56 & 0.05 & 0.45 & 0.56 & 0.65 & 0.56 \\
& k=2    & 0.59 & 0.04 & 0.50 & 0.59 & 0.68 & 0.60 \\
\bottomrule
\end{tabular}
\end{center}
\end{table}

\begin{figure}[!ht]
\centering
  \includegraphics[width=\textwidth]{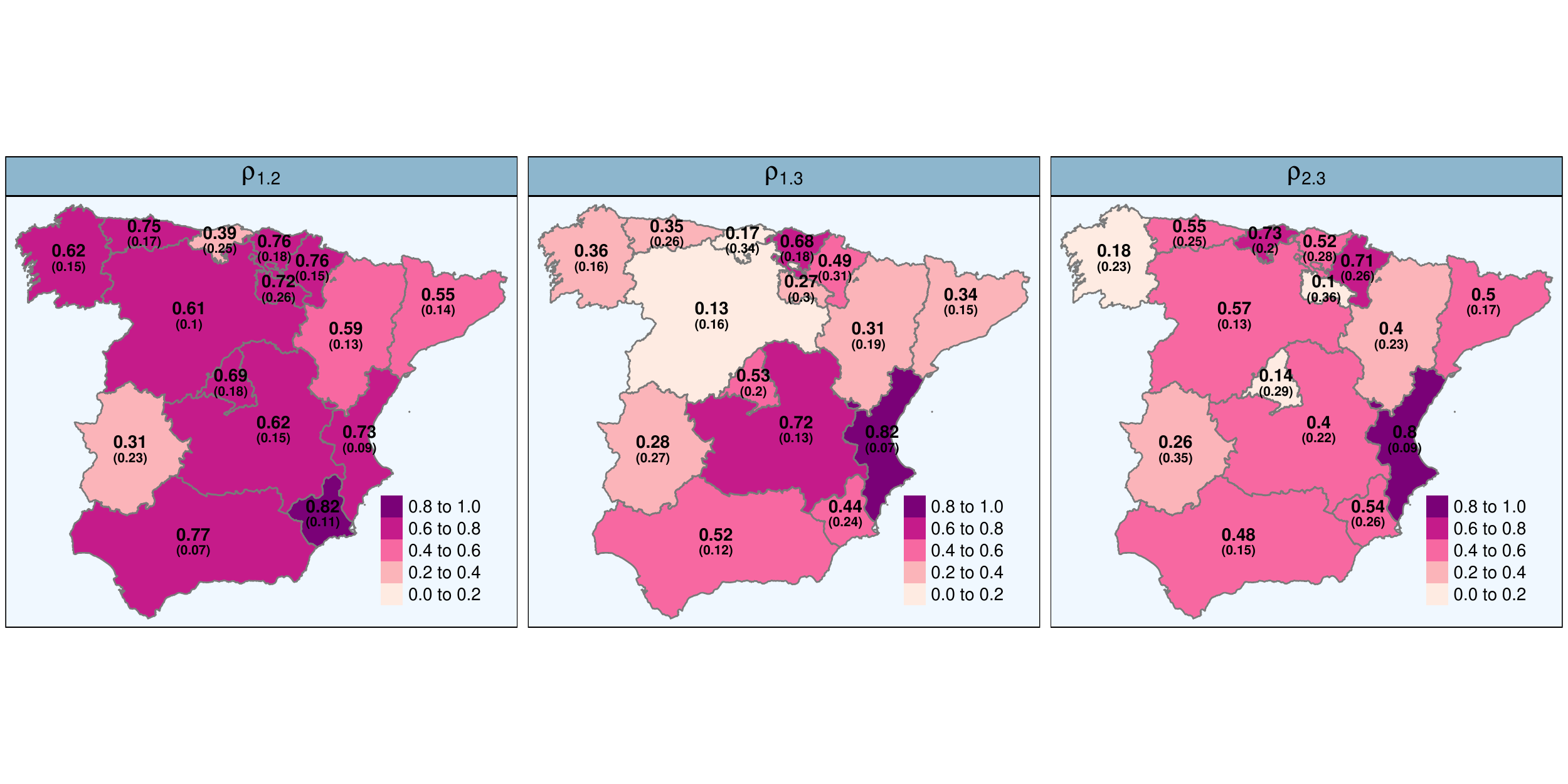}
  \vspace{-2cm}
  \caption{Maps of posterior medians of between-disease correlations and standard deviation (in brackets) for the different subdivisions obtained with the 1st-order neighbourhood partition model. Correlations between lung and colorectal cancer are displayed on the left ($\rho_{1,2}$), the central map displays the correlations between lung and stomach cancer ($\rho_{1,3}$), and the map on the right displays the correlation between colorectal and stomach cancer ($\rho_{2,3}$).}
\label{Figure4}
\end{figure}

Finally, \autoref{Figure4} displays a map with the posterior medians and standard deviations of the between-diseases correlations $\rho_{1,2}$ (left), $\rho_{1,3}$ (center), and $\rho_{2,3}$ (right), for the different subdivisions (Autonomous Regions) obtained with the 1st-order neighbourhood partition model. Not only does the partition model provide correlation for the complete spatial domain (whole Spain), but it also gives correlation for the different subdivisions. This is an advantage over the global models as we add information at different administrative divisions. Moreover, the variability in the posterior medians of the correlations across the subdivisions may indicate a lack of stationarity that the global model cannot cope with, and hence the advantages of the partition models.

\section{Discussion} \label{sec:discussion}
Spatial areal models have a long tradition in epidemiology to study the geographical pattern of a disease. While initially focused on modelling a single disease, spatial models have evolved into a multivariate framework with two notable objectives: to improve estimates by borrowing strength from other diseases, in addition to borrowing information from neighbouring areas, and to estimate latent correlations between the spatial patterns of the diseases under study  to address the connections between them and to hypothesize common risk factors.
Research on spatial multivariate models has received considerable attention in recent years, although their use is not yet widespread in epidemiology mainly because (i) the implementation of multivariate models in available software requires advanced computing skills and (ii) computational issues are accentuated when the number of small areas is large as computing time may become prohibitive.
\cite{vicente2020SERRA} and \cite{vicente2021Psplines} provide an implementation of multivariate CAR and P-splines in R-INLA that can be used by a wide audience without advanced computer skills.

In this paper, we present a new approach to analyse multivariate areal count data when the number of small areas is very large. In particular, we combine the methodology proposed by \cite{orozco2021scalable} for high-dimensional disease mapping with a modification of the multivariate approach given by \cite{botella2015unifying} to avoid overparametrization, obtaining a scalable Bayesian modelling approach to multivariate disease mapping. Our proposal begins with the partition of the spatial domain into subregions with a reduced number of small areas, so that spatial multivariate models can be fitted simultaneously (using both parallel or distributed computation strategies) in each of these regions, reducing computational time and avoiding memory and storage problems.
Dividing the whole spatial domain into disjoint regions may induce border effects as the areas in the limits of a given subdivision do not borrow information from neighbouring areas located in a different subregion. To overcome this issue, we consider $k$-order neighbourhood models that incorporate neighbouring areas to those regions located on the partition boundary.
Finally, variance parameters and between-disease correlations for the whole area are obtained by means of an adaptation of a consensus Monte Carlo algorithm. The correlation coefficients indicate potential geographic factors related (or not) to the different diseases. If the covariance structure is separable, the variance parameters measure the amount of smoothing for each disease.



In addition to the CMC algorithm, we have also considered the Weierstrass rejection sampler (WRS) proposed by \cite{wang2014weierstrass} to recover the parameters of interest for the whole study region (results not shown to save space). In this algorithm, the posterior of the target distribution in the whole area is approximated by combining posterior samples of the subdivisions using rejection sampling. Though it was originally proposed to combine posterior draws from independent MCMC subset chains, it can be adapted to other Bayesian estimation techniques such as INLA through the R package \texttt{weierstrass} (available at https://github.com/wwrechard/weierstrass). In general, very similar posterior marginal estimates are
obtained with both algorithms.

One of the key issues with partition models is to choose the neighbourhood order. Here we use model selection criteria such as DIC and WAIC. Our conclusions are that, in general, the larger the neighbourhood order, the more similar the partition model is to the global model. However, increasing too much the neighbourhood order, the benefits of our proposal in terms of computational time vanish. Overall, first or second order neighbourhood models are appropriate.
%
%
From the simulation study, we conclude that even when the underlying generating process is the Global model, the partition models are very competitive in terms of risk estimation accuracy. Moreover, the global between-disease correlation coefficients are well recovered with the partition models. If the geographical distribution and correlation structure of the underlying process varies across the whole map (which seems very realistic in practice), better results are obtained with our modelling proposals than with the usual global model.
In conclusion, we could argue that partition models have several advantages over a global model. First, they speed up computations and alleviate memory and storage problems. Second, we kill two birds with one stone, as we can provide a global spatial pattern for the whole region and local patterns for the subdivisions, which in our case are of great interest.

In our case study, we use an administrative division of the municipalities of continental Spain corresponding to $D=15$ Autonomous Regions. This partition is a natural choice as Autonomous Regions in Spain are responsible for developing and implementing health policies, and life style may change from region to region. Having estimates using these subdivision may discover associations between diseases that might be associated to specific plans in those regions, the different life styles or other geographical factors having a local influence. This could explain the differences observed in the between-disease correlations in the different subdivisions. On the other hand, this partition may have some inconveniences. For example, the Region of Castilla y Le{\'o}n has 2245 municipalities, still a large number. To overcome this problem, we have also fitted the partition model using a finer subdivision based on 47 provinces rather than on Autonomous Regions. In general results are similar, though the global between-disease correlations are better recovered with the partition based on Autonomous Regions.

The M-models for multivariate disease mapping described in this paper are implemented in the \texttt{R} package \texttt{bigDM}, which also includes several scalable spatial and spatio-temporal Poisson mixed models for high-dimensional areal count data in a fully Bayesian setting using the integrated nested Laplace approximation (INLA) technique. The package also contains a vignette to replicate the data analysis described in Section~\ref{sec:Case_study} using simulated cancer mortality data for the Spanish municipalities, in order to preserve the confidentiality of the original data.

\section*{Acknowledgements}
\addcontentsline{toc}{section}{Acknowledgements}
This work has been supported by the project PID2020-113125RB-I00/MCIN/AEI/10.13039/501100011033. It has also been partially funded by the Public University of Navarra (project PJUPNA2001)


\bibliographystyle{apalike}
\bibliography{refs}   

\begin{thebibliography}{}

\bibitem[Adin et~al., 2022]{bigDM}
Adin, A., Orozco-Acosta, E., and Ugarte, M.~D. (2022).
\newblock {\em bigDM: Scalable Bayesian Disease Mapping Models for
  High-Dimensional Data}.
\newblock R package version 0.5.0.

\bibitem[Besag, 1974]{besag1974spatial}
Besag, J. (1974).
\newblock Spatial interaction and the statistical analysis of lattice systems
  (with discussion).
\newblock {\em Journal of the Royal Statistical Society: Series B (Statistical
  Methodology)}, 36(2):192--225.

\bibitem[Besag et~al., 1991]{besag1991bayesian}
Besag, J., York, J., and Molli{\'e}, A. (1991).
\newblock A {B}ayesian image restoration, with two applications in spatial
  statistics.
\newblock {\em Annals of the Institute of Statistical Mathemathics},
  43(1):1--21.

\bibitem[Botella-Rocamora et~al., 2015]{botella2015unifying}
Botella-Rocamora, P., Martinez-Beneito, M.~A., and Banerjee, S. (2015).
\newblock A unifying modeling framework for highly multivariate disease
  mapping.
\newblock {\em Statistics in Medicine}, 34(9):1548--1559.

\bibitem[Chung et~al., 2015]{Chung2015}
Chung, Y., Gelman, A., Rabe-Hesketh, S., Liu, J., and Dorie, V. (2015).
\newblock Weakly informative prior for point estimation of covariance matrices
  in hierarchical models.
\newblock {\em Journal of Educational and Behavioral Statistics},
  40(2):136--157.

\bibitem[Corpas-Burgos et~al., 2019]{corpas2019convenience}
Corpas-Burgos, F., Botella-Rocamora, P., and Martinez-Beneito, M.~A. (2019).
\newblock On the convenience of heteroscedasticity in highly multivariate
  disease mapping.
\newblock {\em Test}, 28(4):1229--1250.

\bibitem[Cressie and Johannesson, 2008]{cressie2008fixed}
Cressie, N. and Johannesson, G. (2008).
\newblock Fixed rank kriging for very large spatial data sets.
\newblock {\em Journal of the Royal Statistical Society: Series B (Statistical
  Methodology)}, 70(1):209--226.

\bibitem[Dean et~al., 2001]{dean2001detecting}
Dean, C.~B., Ugarte, M.~D., and Militino, A.~F. (2001).
\newblock Detecting interaction between random region and fixed age effects in
  disease mapping.
\newblock {\em Biometrics}, 57(1):197--202.

\bibitem[Eberly and Carlin, 2000]{eberly2000identifiability}
Eberly, L.~E. and Carlin, B.~P. (2000).
\newblock {Identifiability and convergence issues for Markov chain Monte Carlo
  fitting of spatial models}.
\newblock {\em Statistics in Medicine}, 19(17-18):2279--2294.

\bibitem[Fr{\"u}hwirth-Schnatter, 2006]{fruhwirth2006finite}
Fr{\"u}hwirth-Schnatter, S. (2006).
\newblock {\em {Finite mixture and Markov switching models}}.
\newblock Springer Science \& Business Media.

\bibitem[Gelman et~al., 2014]{gelman2014understanding}
Gelman, A., Hwang, J., and Vehtari, A. (2014).
\newblock {Understanding predictive information criteria for Bayesian models}.
\newblock {\em Statistics and Computing}, 24(6):997--1016.

\bibitem[Goicoa et~al., 2018]{goicoa2018spatio}
Goicoa, T., Adin, A., Ugarte, M.~D., and Hodges, J.~S. (2018).
\newblock In spatio-temporal disease mapping models, identifiability
  constraints affect {PQL} and {INLA} results.
\newblock {\em Stochastic Environmental Research and Risk Assessment},
  32(3):749--770.

\bibitem[Goicoa et~al., 2012]{goicoa2012comparing}
Goicoa, T., Ugarte, M., Etxeberria, J., and Militino, A. (2012).
\newblock Comparing {CAR} and {P}-spline models in spatial disease mapping.
\newblock {\em Environmental and Ecological Statistics}, 19(4):573--599.

\bibitem[Jin et~al., 2007]{jin2007order}
Jin, X., Banerjee, S., and Carlin, B. (2007).
\newblock Order-free co-regionalized areal data models with application to
  multiple-disease mapping.
\newblock {\em Journal of the Royal Statistical Society: Series B (Statistical
  Methodology)}, 69(5):817--838.

\bibitem[Katzfuss, 2017]{katzfuss2017Mult}
Katzfuss, M. (2017).
\newblock A multi-resolution approximation for massive spatial datasets.
\newblock {\em Journal of the American Statistical Association},
  112(517):201--214.

\bibitem[Katzfuss and Guinness, 2021]{katzfuss2021general}
Katzfuss, M. and Guinness, J. (2021).
\newblock {A general framework for Vecchia approximations of Gaussian
  processes}.
\newblock {\em Statistical Science}, 36(1):124--141.

\bibitem[Leroux et~al., 1999]{leroux1999estimation}
Leroux, B.~G., Lei, X., and Breslow, N. (1999).
\newblock Estimation of disease rates in small areas: a new mixed model for
  spatial dependence.
\newblock {\em In , Halloran, M. Berry, D. (eds). Statistical Models in
  Epidemiology, the Environment, and Clinical Trials}, pages 179--192.

\bibitem[Li et~al., 2014]{LiRich2014}
Li, G., Haining, R., Richardson, S., and Best, N. (2014).
\newblock Space-time variability in burglary risk: A {B}ayesian spatio-temporal
  modelling approach.
\newblock {\em Spatial Statistics}, 9:180--191.

\bibitem[Lindgren and Rue, 2015]{LindRue2015}
Lindgren, F. and Rue, H. (2015).
\newblock {Bayesian spatial modelling with R-INLA}.
\newblock {\em Journal of Statistical Software}, 63:1--25.

\bibitem[Lindgren et~al., 2011]{lindgren2011explicit}
Lindgren, F., Rue, H., and Lindstr{\"o}m, J. (2011).
\newblock {An explicit link between Gaussian fields and Gaussian Markov random
  fields: the stochastic partial differential equation approach}.
\newblock {\em Journal of the Royal Statistical Society: Series B (Statistical
  Methodology)}, 73(4):423--498.

\bibitem[Lindsay, 1995]{lindsay1995mixture}
Lindsay, B.~G. (1995).
\newblock Mixture models: theory, geometry, and applications.
\newblock In: NSF-CBMS Regional Conference Series in Probability and
  Statistics, JSTOR.

\bibitem[MacNab, 2016]{macnab2016lineara}
MacNab, Y.~C. (2016).
\newblock Linear models of coregionalization for multivariate lattice data: a
  general framework for coregionalized multivariate {CAR} models.
\newblock {\em Statistics in Medicine}, 35(21):3827--3850.

\bibitem[MacNab, 2018]{MacNab2018Test}
MacNab, Y.~C. (2018).
\newblock Some recent work on multivariate {G}aussian {M}arkov random fields.
\newblock {\em Test}, 27(3):497--541.

\bibitem[MacNab, 2022]{MacNab2022SpatStat}
MacNab, Y.~C. (2022).
\newblock {Bayesian disease mapping: Past, present, and future}.
\newblock {\em Spatial Statistics}, 50:100593.

\bibitem[Mardia, 1988]{mardia1988multi}
Mardia, K. (1988).
\newblock Multi-dimensional multivariate {G}aussian {M}arkov random fields with
  application to image processing.
\newblock {\em Journal of Multivariate Analysis}, 24(2):265--284.

\bibitem[Martinez-Beneito, 2013]{martinez2013general}
Martinez-Beneito, M.~A. (2013).
\newblock A general modelling framework for multivariate disease mapping.
\newblock {\em Biometrika}, 100(3):539--553.

\bibitem[Nychka et~al., 2015]{nychka2015multiresolution}
Nychka, D., Bandyopadhyay, S., Hammerling, D., Lindgren, F., and Sain, S.
  (2015).
\newblock {A multiresolution Gaussian process model for the analysis of large
  spatial datasets}.
\newblock {\em Journal of Computational and Graphical Statistics},
  24(2):579--599.

\bibitem[Orozco-Acosta et~al., 2021]{orozco2021scalable}
Orozco-Acosta, E., Adin, A., and Ugarte, M.~D. (2021).
\newblock {Scalable Bayesian modelling for smoothing disease risks in large
  spatial data sets using INLA}.
\newblock {\em Spatial Statistics}, 41:100496.

\bibitem[Orozco-Acosta et~al., 2022]{orozco2022}
Orozco-Acosta, E., Adin, A., and Ugarte, M.~D. (2022).
\newblock {Big problems in spatio-temporal disease mapping: methods and
  software}.
\newblock {\em arXiv:2201.08323}.

\bibitem[Pe{\~n}a and Irie, 2022]{pena2022}
Pe{\~n}a, V. and Irie, K. (2022).
\newblock {On the relationship between Uhlig extended and beta-Bartlett
  processes}.
\newblock {\em Journal of Time Series Analysis}, 43(1):147--153.

\bibitem[Pettit, 1990]{pettit1990conditional}
Pettit, L. (1990).
\newblock The conditional predictive ordinate for the normal distribution.
\newblock {\em Journal of the Royal Statistical Society: Series B
  (Methodological)}, 52(1):175--184.

\bibitem[Riebler et~al., 2016]{riebler2016intuitive}
Riebler, A., S{\o}rbye, S.~H., Simpson, D., and Rue, H. (2016).
\newblock An intuitive bayesian spatial model for disease mapping that accounts
  for scaling.
\newblock {\em Statistical Methods in Medical Research}, 25(4):1145--1165.

\bibitem[Rue et~al., 2009]{rue2009approximate}
Rue, H., Martino, S., and Chopin, N. (2009).
\newblock Approximate {B}ayesian inference for latent {G}aussian models by
  using integrated nested {L}aplace approximations.
\newblock {\em Journal of the Royal Statistical Society: Series B (Statistical
  Methodology)}, 71(2):319--392.

\bibitem[Sain et~al., 2011]{sain2011}
Sain, S.~R., Furrer, R., and Cressie, N. (2011).
\newblock A spatial analysis of multivariate output from regional climate
  models.
\newblock {\em The Annals of Applied Statistics}, 5(1):150--175.

\bibitem[Scott et~al., 2016]{scott2016consensus}
Scott, S.~L., Blocker, A.~W., Bonassi, F.~V., Chipman, H.~A., George, E.~I.,
  and McCulloch, R.~E. (2016).
\newblock {Bayes and big data: The consensus Monte Carlo algorithm}.
\newblock {\em International Journal of Management Science and Engineering
  Management}, 11(2):78--88.

\bibitem[Spiegelhalter et~al., 2002]{spiegelhalter2002bayesian}
Spiegelhalter, D.~J., Best, N.~G., Carlin, B.~P., and Van Der~Linde, A. (2002).
\newblock Bayesian measures of model complexity and fit.
\newblock {\em Journal of the Royal Statistical Society: Series B (Statistical
  Methodology)}, 64(4):583--639.

\bibitem[Ugarte et~al., 2017]{ugarte2017one}
Ugarte, M.~D., Adin, A., and Goicoa, T. (2017).
\newblock {One-dimensional, two-dimensional, and three dimensional B-splines to
  specify space-time interactions in Bayesian disease mapping: Model fitting
  and model identifiability}.
\newblock {\em Spatial Statistics}, 22:451--468.

\bibitem[Ugarte et~al., 2010]{ugarte2010spatio}
Ugarte, M.~D., Goicoa, T., and Militino, A.~F. (2010).
\newblock Spatio-temporal modeling of mortality risks using penalized splines.
\newblock {\em Environmetrics}, 21(3-4):270--289.

\bibitem[Vicente et~al., 2020a]{vicente2020dowry}
Vicente, G., Goicoa, T., Fern\'andez-Rasines, P., and Ugarte, M.~D. (2020a).
\newblock Crime against women in {I}ndia: unveiling spatial patterns and
  temporal trends of dowry deaths in the districts of {U}ttar {P}radesh.
\newblock {\em Journal of the Royal Statistical Society: Series A (Statistics
  in Society)}, 183(2):655--679.

\bibitem[Vicente et~al., 2018]{vicente2018small}
Vicente, G., Goicoa, T., Puranik, A., and Ugarte, M.~D. (2018).
\newblock Small area estimation of gender-based violence: rape incidence risks
  in {U}ttar {P}radesh, {I}ndia.
\newblock {\em Statistics and Applications}, 16(1):71--90.

\bibitem[Vicente et~al., 2020b]{vicente2020SERRA}
Vicente, G., Goicoa, T., and Ugarte, M.~D. (2020b).
\newblock {Bayesian inference in multivariate spatio-temporal areal models
  using INLA: analysis of gender-based violence in small areas}.
\newblock {\em Stochastic Environmental Research and Risk Assessment},
  34(10):1421--1440.

\bibitem[Vicente et~al., 2021]{vicente2021Psplines}
Vicente, G., Goicoa, T., and Ugarte, M.~D. (2021).
\newblock {Multivariate Bayesian spatio-temporal P-spline models to analyze
  crimes against women}.
\newblock {\em Biostatistics (in press).
  https://doi.org/10.1093/biostatistics/kxab042}.

\bibitem[Wang and Dunson, 2014]{wang2014weierstrass}
Wang, X. and Dunson, D.~B. (2014).
\newblock {Parallelizing MCMC via Weierstrass sampler}.

\bibitem[Watanabe, 2010]{watanabe2010asymptotic}
Watanabe, S. (2010).
\newblock Asymptotic equivalence of {B}ayes cross validation and widely
  applicable information criterion in singular learning theory.
\newblock {\em Journal of Machine Learning Research}, 11(Dec):3571--3594.

\end{thebibliography}

\newpage

\setcounter{section}{0} 
\renewcommand{\thesection}{\Alph{section}}

\setcounter{figure}{0}
\renewcommand\thefigure{\thesection.\arabic{figure}}

\setcounter{table}{0}
\renewcommand\thetable{\thesection.\arabic{table}}

\section{Appendix}  \label{sec:AppendixA}

In this Appendix we briefly explain how to implement the Bartlett decomposition in R-INLA. This requires that the hyperparameters have support on $\mathbb{R}$. So, we will reparameterise the elements $c_j$ described in Equation~\eqref{Cholesky_factor} as
\begin{equation*}
	\theta_{j}=\log(c_j),\: j=1,\ldots,J,
\end{equation*}
and the log priors for $c_j$ are given as the corresponding log priors for $\theta_{j}, \forall j=1,\ldots,J$.\\

\noindent
For each $c^2_j$, $\forall j=1,\ldots,J$, we assign a chi-square distribution with $J+2-j+1$ degrees of freedom, so the log prior for $\theta_{j}$ is
\begin{equation*}
	\log \pi(\theta_j) = \log(2) + 2\cdot \theta_j + \log f_j\left[\exp(2\theta_j) \right]
\end{equation*}
where $f_j(\cdot)$ is the probability density function (pdf) of $c^2_j$. This expression is obtained as follows.
	\begin{eqnarray*}
		\theta_j &=& \log(c_j) = \frac{1}{2}\log(c^2_j) = \frac{1}{2}\log(x_j) \Rightarrow
		x_j = g^{-1}(\theta_j) = \exp (2\theta_j) \\
		\frac{d x_j}{d\theta_j} &=& 2 \exp (2\theta_j) \Rightarrow  \left| \frac{d x_j}{d\theta_j} \right| = 2 \exp (2\theta_j) \\
	\pi(\theta_j)
			&=& f_{j}\left[ g^{-1}(\theta_j) \right] \left| \frac{d x_j}{d\theta_j} \right|
			= f_{j}\left[ \exp (2\theta_j)  \right] \: 2 \exp (2\theta_j)  \\
	\log \pi(\theta_j) &=& \log f_{j}\left[\exp(2\theta_j) \right] + \log(2) + 2\cdot \theta_j
	\end{eqnarray*}

\noindent
Note that non-diagonal elements in $\mathbf{A}$ (see Equation~\eqref{Cholesky_factor}) have support on $\mathbb{R}$, so there is no need to  reparameterize them, i.e.,
\begin{equation*}
	\theta_{j}=n_{il},\: j=J+1,\ldots,J(J+1)/2.
\end{equation*}

\noindent
Finally, let us denote $\ttheta=(\theta_1,\ldots,\theta_J,\theta_{J+1},\ldots,\theta_{J(J+1)/2})'$.
Then,
\begin{equation*}
\pi(\ttheta)
		= \prod_{j=1}^{J(J+1)/2} \pi(\theta_j)
		= \prod_{j=1}^{J} \pi(\theta_j) \times \prod_{j=J+1}^{J(J+1)/2} \pi(\theta_j),
\end{equation*}
and taking logarithms
%
	\begin{eqnarray*}
		\log \pi(\ttheta)
		&=& \sum_{j=1}^{J} \log\pi(\theta_j) + \sum_{j=J+1}^{J(J+1)/2} \log\pi(\theta_j)	\\
		&=& \sum_{j=1}^{J} \left\lbrace \log(2) + 2\cdot \theta_j + \log f_j\left[\exp(2\theta_j) \right] \right\rbrace
		+ \sum_{j=J+1}^{J(J+1)/2} \log \phi(\theta_j) \\
		&=& J\log(2) + 2 \sum_{j=1}^{J} \theta_j + \sum_{j=1}^{J} \log f_j\left[\exp(2\theta_j) \right]
		+ \sum_{j=J+1}^{J(J+1)/2} \log \phi(\theta_j)
	\end{eqnarray*}
%
where $f_j(\cdot)$ are the pdf of the chi-squared distribution with $J+2-j+1$ degrees of freedom, $j=1,\ldots,J$, and $\phi(\cdot)$ is the pdf of the standard normal distribution.

\subsection*{Code} \label{sec:code}
The R-INLA code to assign log prior distributions to the hyperparameters of the M-models (elements of the $\mathbf{A}$ matrix) can be checked in the \texttt{Mmodel\_icar()} function of the \texttt{bigDM} package.

\section{Appendix}  \label{sec:AppendixB}
In this Appendix we include additional tables and figures regarding the simulation study (Section~\ref{sec:simulation}) and the results of the joint analysis of mortality data for lung, colorectal and stomach cancer (case study of Section~\ref{sec:Case_study}).

\begin{figure}[!ht]
\begin{center}
    \includegraphics[width=0.85\textwidth]{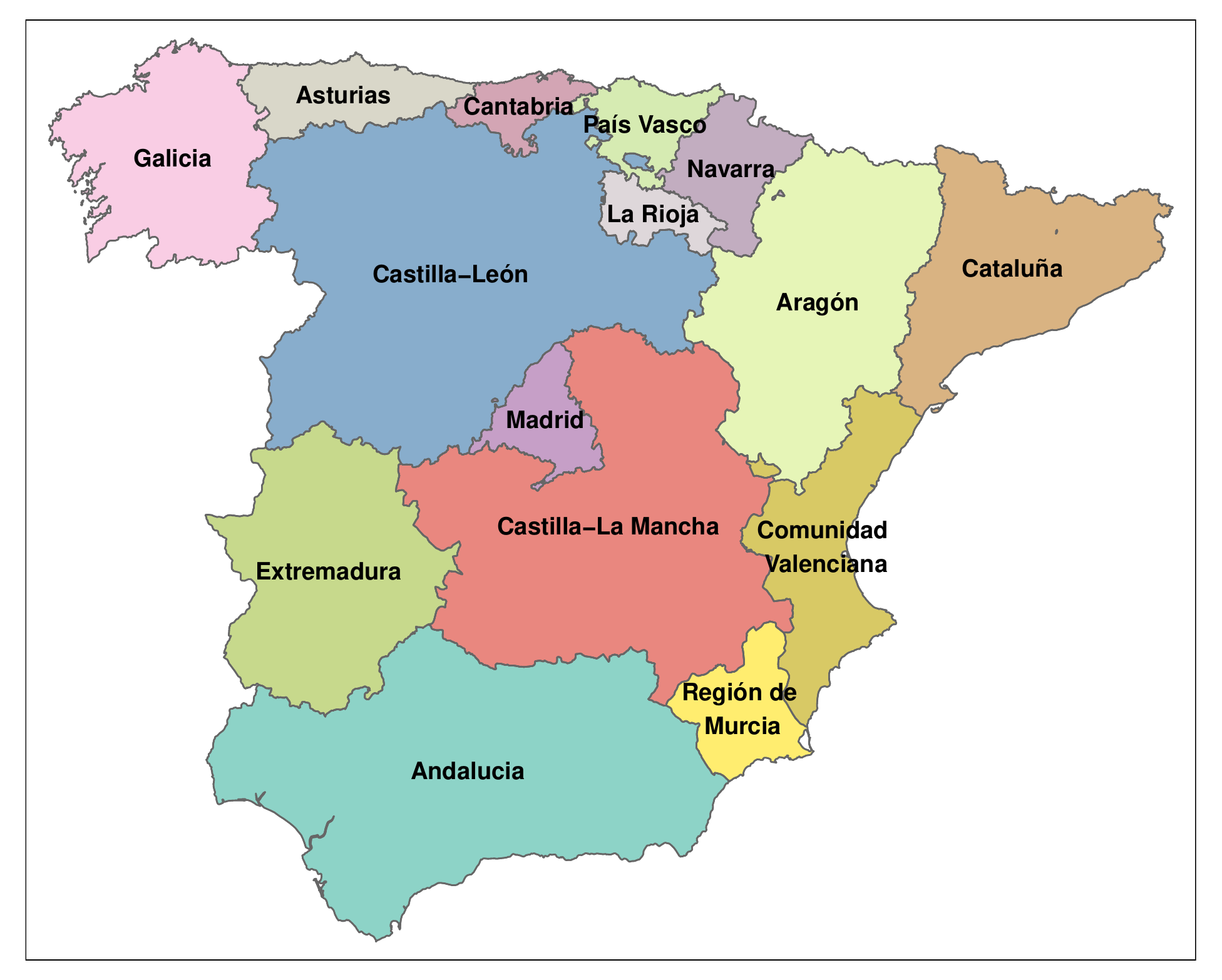}\hfill
\end{center}
\caption{Map of the administrative division of Spain into Autonomous Regions.}
\label{fig:Map_CCAA}
\end{figure}

\autoref{fig:Map_CCAA} displays the map of the administrative division of Spain into Autonomous Regions.

\autoref{tab:Scenario2_k0}, \autoref{tab:Scenario2_k1} and \autoref{tab:Scenario2_k2} compares the true values of model parameters (local correlation coefficients in each subdivision) against average values of posterior mean estimates over the 100 simulated data sets for Scenario 2.

\begin{figure}[!ht]
\begin{center}
    \vspace{-1.5cm}
    \includegraphics[width=1.25\textwidth]{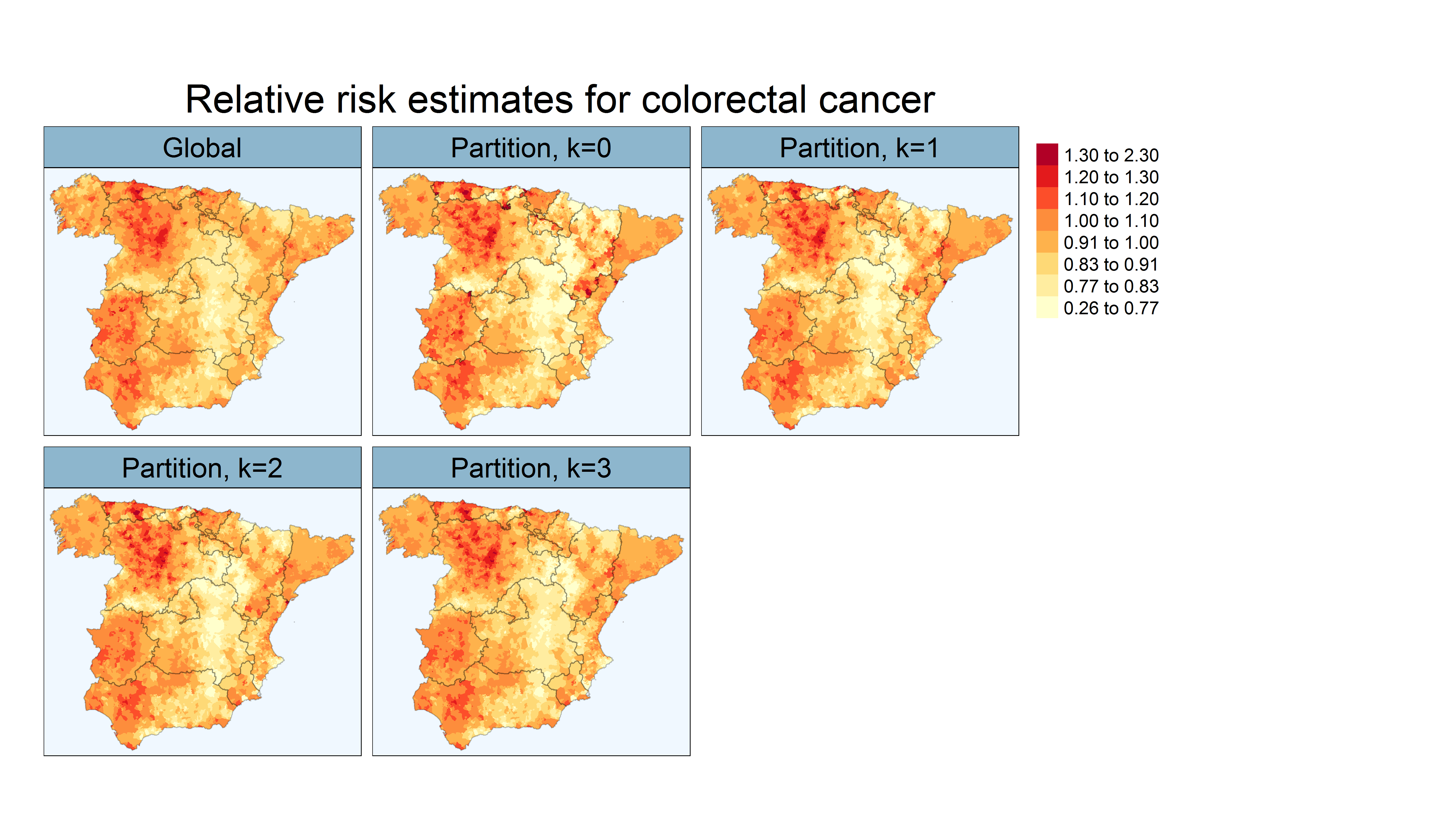}\hfill
    \vspace{-1.5cm}
    \includegraphics[width=1.25\textwidth]{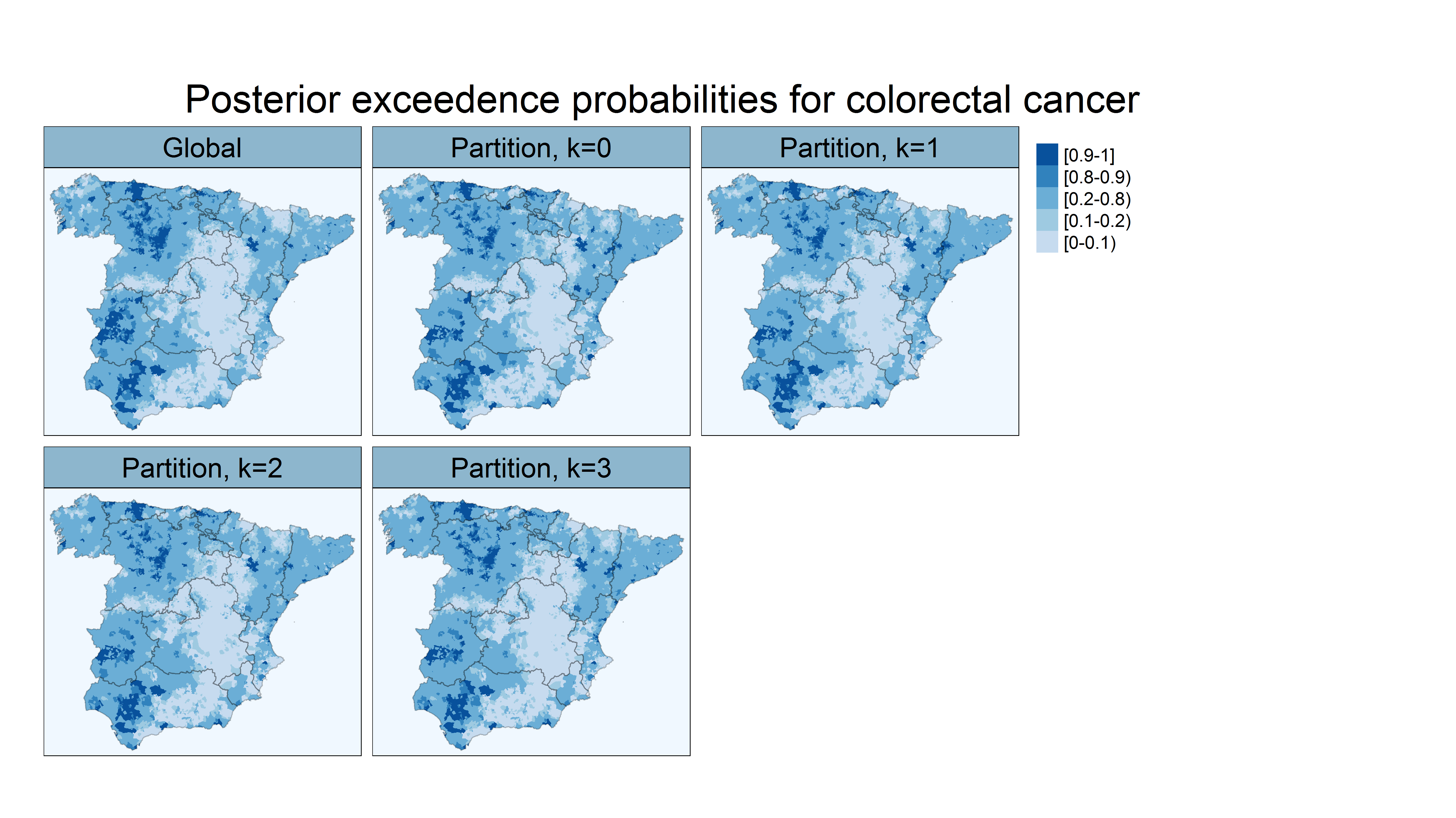}\hfill
    \vspace{-1.5cm}
\end{center}
\caption{Maps of posterior median estimates of mortality relative risk for colorectal cancer (top) and posterior exceedance probabilities $P(R_{ij}>1 \vert \mathbf{O})$ (bottom) in continental Spain.}
\label{fig:risk_icar_colorectal}
\end{figure}

\begin{figure}[!ht]
\begin{center}
    \vspace{-1.5cm}
    \includegraphics[width=1.25\textwidth]{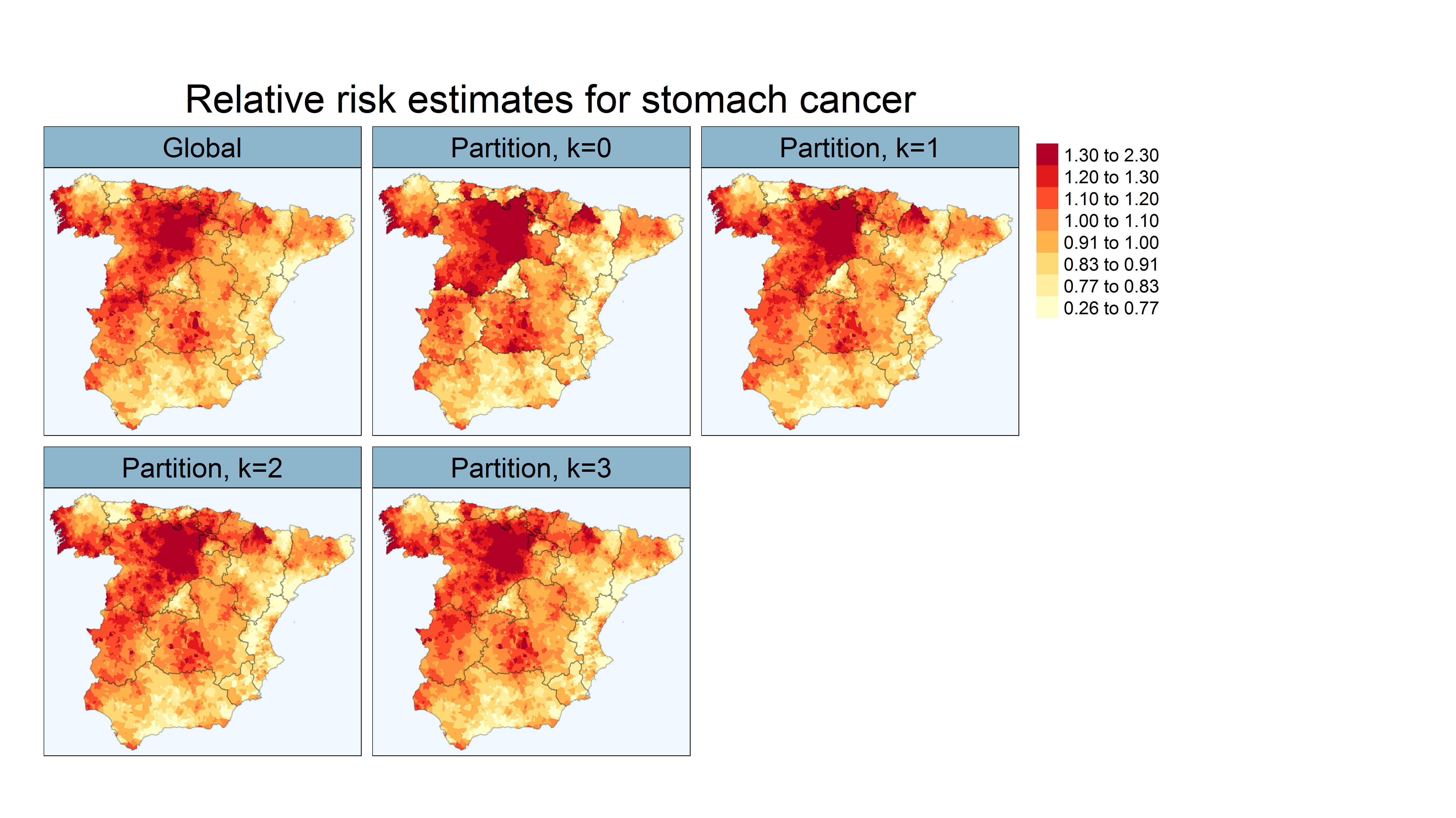}\hfill
    \vspace{-1.5cm}
    \includegraphics[width=1.25\textwidth]{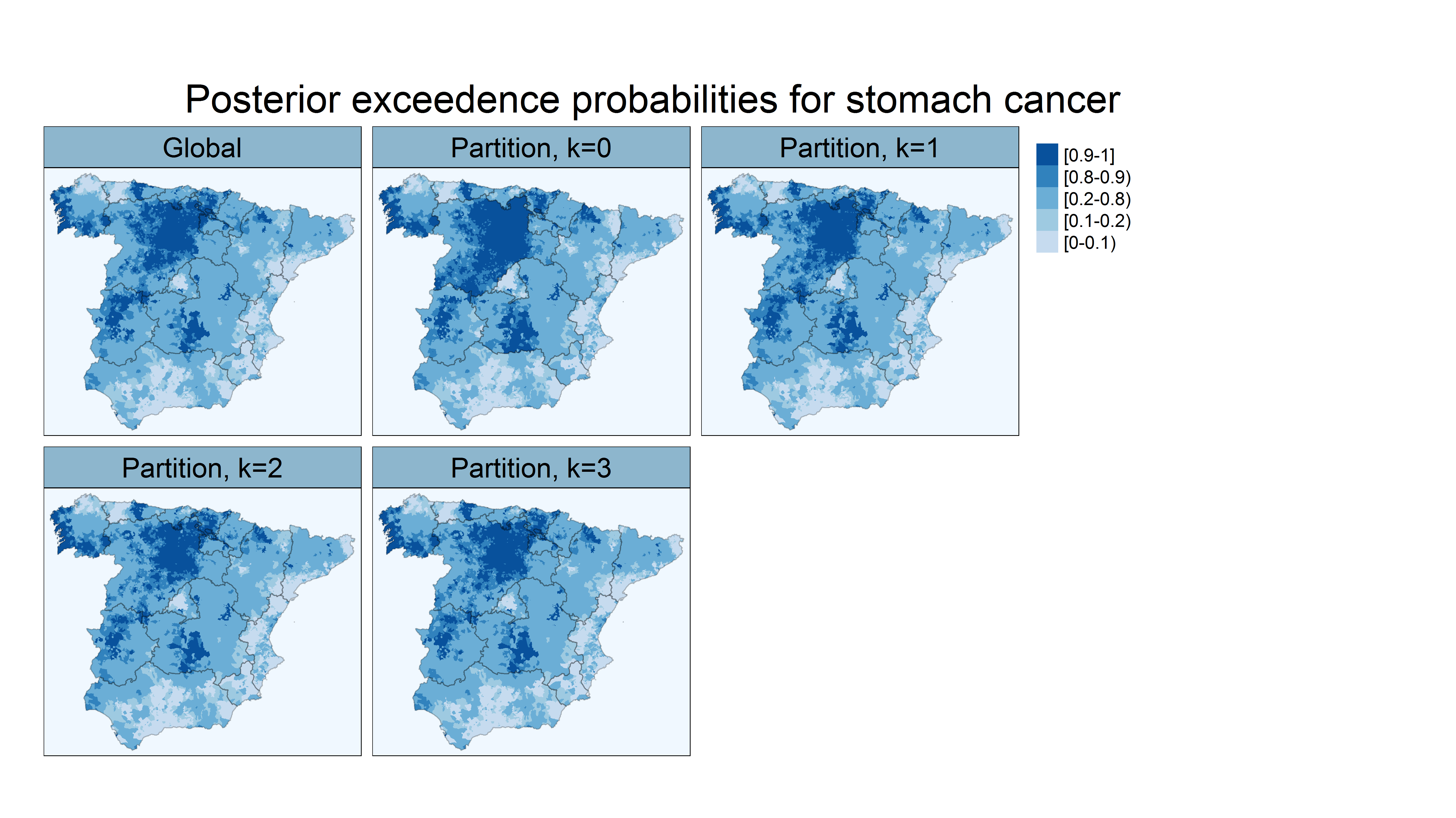}\hfill
    \vspace{-1.5cm}
\end{center}
\caption{Maps of posterior median estimates of mortality relative risk for stomach cancer (top) and posterior exceedance probabilities $P(R_{ij}>1 \vert \mathbf{O})$ (bottom) in continental Spain.}
\label{fig:risk_icar_stomach}
\end{figure}

\autoref{fig:risk_icar_colorectal} displays the spatial patterns of colorectal cancer mortality risks (top) and the posterior probabilities of risk exceedance (bottom), $P(R_{ij}>1 \vert \mathbf{O})$, obtained with the global and the disjoint models. Similarly, \autoref{fig:risk_icar_stomach} displays the spatial patterns of stomach cancer mortality risks (top) and the posterior probabilities of risk exceedance (bottom), $P(R_{ij}>1 \vert \mathbf{O})$, obtained with the global and the disjoint models.



\clearpage
\begin{landscape}
\begin{table*}[!ht]
\caption{{\bf Disjoint model}. Average values of posterior mean, posterior standard deviation (SD), simulated standard errors (sim) and empirical coverage of the 95\% credible intervals (Cov) for local estimates model parameters based on 100 simulated data sets for Scenario 2.}
\label{tab:Scenario2_k0}
\vspace{-0.5cm}
\renewcommand{\arraystretch}{1.1}
\begin{center}
\resizebox{1.4\textwidth}{!}{
\begin{tabular}{c|crrrr|crrrr|crrrr|crrrr|crrrr}
\toprule
& \multicolumn{5}{c|}{\bf Andaluc\'ia} & \multicolumn{5}{c|}{\bf Arag\'on} & \multicolumn{5}{c|}{\bf Asturias} & \multicolumn{5}{c|}{\bf Cantabria} & \multicolumn{5}{c}{\bf Castilla - La Mancha}\\
& & & & & & & & & & & & & & & & & & & & & \\
Parameter & Value & Mean & SD & Sim & Cov & Value & Mean & SD & Sim & Cov & Value & Mean & SD & Sim & Cov & Value & Mean & SD & Sim & Cov & Value & Mean & SD & Sim & Cov \\
\hline
$\alpha_1$   & -0.20 & -0.20 & 0.01 & 0.01 & 0.97 & -0.20 & -0.21 & 0.03 & 0.03 & 0.91 & -0.20 & -0.20 & 0.03 & 0.04 & 0.95 & -0.20 & -0.20 & 0.04 & 0.04 & 0.96 & -0.20 & -0.20 & 0.02 & 0.02 & 0.88 \\
$\alpha_2$   & -0.10 & -0.10 & 0.02 & 0.01 & 0.95 & -0.10 & -0.10 & 0.03 & 0.03 & 0.98 & -0.10 & -0.10 & 0.04 & 0.04 & 0.96 & -0.10 & -0.11 & 0.05 & 0.05 & 0.93 & -0.10 & -0.10 & 0.02 & 0.02 & 0.96 \\
$\alpha_3$   &  0.10 &  0.10 & 0.02 & 0.02 & 0.97 &  0.10 &  0.09 & 0.05 & 0.04 & 0.97 &  0.10 &  0.09 & 0.05 & 0.05 & 0.94 &  0.10 &  0.08 & 0.07 & 0.07 & 0.94 &  0.10 &  0.09 & 0.03 & 0.03 & 0.88 \\
$\sigma^2_1$ &  0.50 &  0.51 & 0.05 & 0.04 & 0.96 &  0.50 &  0.53 & 0.09 & 0.08 & 0.93 &  0.50 &  0.58 & 0.15 & 0.14 & 0.94 &  0.50 &  0.59 & 0.16 & 0.14 & 0.95 &  0.50 &  0.50 & 0.05 & 0.06 & 0.91 \\
$\sigma^2_2$ &  0.40 &  0.41 & 0.04 & 0.05 & 0.95 &  0.40 &  0.45 & 0.10 & 0.09 & 0.95 &  0.40 &  0.49 & 0.14 & 0.13 & 0.94 &  0.40 &  0.51 & 0.16 & 0.14 & 0.93 &  0.40 &  0.42 & 0.06 & 0.06 & 0.94 \\
$\sigma^2_3$ &  0.30 &  0.32 & 0.05 & 0.05 & 0.93 &  0.30 &  0.39 & 0.12 & 0.12 & 0.87 &  0.30 &  0.41 & 0.15 & 0.11 & 0.94 &  0.30 &  0.49 & 0.19 & 0.17 & 0.91 &  0.30 &  0.33 & 0.07 & 0.07 & 0.95 \\[1.ex]
$\rho_{12}$  &  0.76 &  0.75 & 0.04 & 0.04 & 0.94 &  0.58 &  0.52 & 0.12 & 0.11 & 0.94 &  0.71 &  0.64 & 0.12 & 0.11 & 0.95 &  0.37 &  0.29 & 0.19 & 0.18 & 0.92 &  0.60 &  0.58 & 0.07 & 0.08 & 0.93 \\
$\rho_{13}$  &  0.52 &  0.52 & 0.07 & 0.08 & 0.94 &  0.30 &  0.24 & 0.17 & 0.20 & 0.90 &  0.34 &  0.28 & 0.20 & 0.19 & 0.96 &  0.14 &  0.11 & 0.23 & 0.20 & 0.96 &  0.71 &  0.67 & 0.08 & 0.07 & 0.99 \\
$\rho_{23}$  &  0.47 &  0.46 & 0.08 & 0.08 & 0.96 &  0.37 &  0.29 & 0.19 & 0.18 & 0.94 &  0.51 &  0.44 & 0.19 & 0.18 & 0.97 &  0.69 &  0.58 & 0.19 & 0.15 & 0.99 &  0.38 &  0.37 & 0.11 & 0.11 & 0.93 \\[1.ex]
\hline
& & & & & & & & & & & & & & & & & & & & & \\[-2.ex]
& \multicolumn{5}{c|}{\bf Castilla y Le\'on} & \multicolumn{5}{c|}{\bf Catalu\~na} & \multicolumn{5}{c|}{\bf Comunidad Valenciana} & \multicolumn{5}{c|}{\bf Extremadura} & \multicolumn{5}{c}{\bf Galicia} \\
& & & & & & & & & & & & & & & & & & & & & \\
Parameter & Value & Mean & SD & Sim & Cov & Value & Mean & SD & Sim & Cov & Value & Mean & SD & Sim & Cov & Value & Mean & SD & Sim & Cov & Value & Mean & SD & Sim & Cov \\
\hline
$\alpha_1$   & -0.20 & -0.20 & 0.02 & 0.02 & 0.96 & -0.20 & -0.20 & 0.02 & 0.02 & 0.94 & -0.20 & -0.20 & 0.02 & 0.02 & 0.96 & -0.20 & -0.20 & 0.02 & 0.02 & 0.96 & -0.20 & -0.20 & 0.01 & 0.02 & 0.95 \\
$\alpha_2$   & -0.10 & -0.11 & 0.03 & 0.03 & 0.93 & -0.10 & -0.10 & 0.02 & 0.02 & 0.96 & -0.10 & -0.10 & 0.02 & 0.02 & 0.94 & -0.10 & -0.10 & 0.03 & 0.03 & 0.96 & -0.10 & -0.10 & 0.02 & 0.02 & 0.96 \\
$\alpha_3$   &  0.10 &  0.10 & 0.04 & 0.03 & 0.97 &  0.10 &  0.10 & 0.03 & 0.03 & 0.93 &  0.10 &  0.10 & 0.03 & 0.03 & 0.92 &  0.10 &  0.09 & 0.04 & 0.04 & 0.95 &  0.10 &  0.10 & 0.03 & 0.03 & 0.93 \\
$\sigma^2_1$ &  0.50 &  0.51 & 0.06 & 0.06 & 0.97 &  0.50 &  0.51 & 0.05 & 0.05 & 0.93 &  0.50 &  0.51 & 0.06 & 0.06 & 0.92 &  0.50 &  0.54 & 0.09 & 0.07 & 0.94 &  0.50 &  0.52 & 0.06 & 0.06 & 0.96 \\
$\sigma^2_2$ &  0.40 &  0.42 & 0.06 & 0.07 & 0.92 &  0.40 &  0.42 & 0.05 & 0.05 & 0.95 &  0.40 &  0.42 & 0.06 & 0.06 & 0.97 &  0.40 &  0.44 & 0.09 & 0.09 & 0.95 &  0.40 &  0.41 & 0.06 & 0.06 & 0.94 \\
$\sigma^2_3$ &  0.30 &  0.34 & 0.08 & 0.08 & 0.91 &  0.30 &  0.31 & 0.05 & 0.05 & 0.95 &  0.30 &  0.32 & 0.06 & 0.06 & 0.93 &  0.30 &  0.36 & 0.10 & 0.10 & 0.93 &  0.30 &  0.32 & 0.06 & 0.07 & 0.89 \\[1.ex]
$\rho_{12}$  &  0.60 &  0.58 & 0.08 & 0.07 & 0.97 &  0.54 &  0.52 & 0.06 & 0.07 & 0.88 &  0.72 &  0.71 & 0.05 & 0.05 & 0.98 &  0.30 &  0.28 & 0.12 & 0.11 & 0.97 &  0.61 &  0.61 & 0.07 & 0.07 & 0.93 \\
$\rho_{13}$  &  0.12 &  0.13 & 0.13 & 0.13 & 0.93 &  0.34 &  0.34 & 0.09 & 0.09 & 0.94 &  0.81 &  0.79 & 0.06 & 0.06 & 0.96 &  0.27 &  0.25 & 0.16 & 0.15 & 0.98 &  0.36 &  0.34 & 0.11 & 0.11 & 0.93 \\
$\rho_{23}$  &  0.56 &  0.53 & 0.11 & 0.11 & 0.96 &  0.48 &  0.46 & 0.09 & 0.09 & 0.96 &  0.79 &  0.76 & 0.07 & 0.07 & 0.93 &  0.24 &  0.22 & 0.18 & 0.17 & 0.92 &  0.17 &  0.16 & 0.12 & 0.12 & 0.96 \\[1.ex]
\hline
& & & & & & & & & & & & & & & & & & & & & \\[-2.ex]
& \multicolumn{5}{c|}{\bf La Rioja} & \multicolumn{5}{c}{\bf Madrid} & \multicolumn{5}{c|}{\bf Murcia} & \multicolumn{5}{c|}{\bf Navarra} & \multicolumn{5}{c}{\bf Pa\'is Vasco} \\
& & & & & & & & & & & & & & & & & & & & & \\
Parameter & Value & Mean & SD & Sim & Cov & Value & Mean & SD & Sim & Cov & Value & Mean & SD & Sim & Cov & Value & Mean & SD & Sim & Cov & Value & Mean & SD & Sim & Cov \\
\hline
$\alpha_1$   & -0.20 & -0.22 & 0.06 & 0.06 & 0.95 & -0.20 & -0.20 & 0.04 & 0.04 & 0.97 & -0.20 & -0.20 & 0.03 & 0.03 & 0.97 & -0.20 & -0.21 & 0.04 & 0.04 & 0.93 & -0.20 & -0.20 & 0.03 & 0.03 & 0.94 \\
$\alpha_2$   & -0.10 & -0.13 & 0.08 & 0.08 & 0.92 & -0.10 & -0.11 & 0.04 & 0.05 & 0.94 & -0.10 & -0.11 & 0.04 & 0.04 & 0.96 & -0.10 & -0.11 & 0.05 & 0.05 & 0.98 & -0.10 & -0.10 & 0.03 & 0.03 & 0.94 \\
$\alpha_3$   &  0.10 &  0.07 & 0.10 & 0.09 & 0.97 &  0.10 &  0.09 & 0.06 & 0.06 & 0.96 &  0.10 &  0.09 & 0.05 & 0.05 & 0.96 &  0.10 &  0.07 & 0.07 & 0.06 & 0.94 &  0.10 &  0.09 & 0.04 & 0.04 & 0.98 \\
$\sigma^2_1$ &  0.50 &  0.70 & 0.22 & 0.16 & 0.90 &  0.50 &  0.55 & 0.10 & 0.09 & 0.96 &  0.50 &  0.64 & 0.18 & 0.17 & 0.91 &  0.50 &  0.57 & 0.13 & 0.14 & 0.89 &  0.50 &  0.53 & 0.09 & 0.09 & 0.94 \\
$\sigma^2_2$ &  0.40 &  0.64 & 0.25 & 0.22 & 0.86 &  0.40 &  0.44 & 0.10 & 0.09 & 0.96 &  0.40 &  0.53 & 0.16 & 0.14 & 0.90 &  0.40 &  0.47 & 0.13 & 0.10 & 0.98 &  0.40 &  0.44 & 0.09 & 0.09 & 0.93 \\
$\sigma^2_3$ &  0.30 &  0.66 & 0.32 & 0.26 & 0.84 &  0.30 &  0.37 & 0.10 & 0.10 & 0.93 &  0.30 &  0.41 & 0.15 & 0.13 & 0.96 &  0.30 &  0.47 & 0.17 & 0.15 & 0.90 &  0.30 &  0.36 & 0.09 & 0.10 & 0.92 \\[1.ex]
$\rho_{12}$  &  0.65 &  0.51 & 0.21 & 0.16 & 0.98 &  0.66 &  0.61 & 0.10 & 0.11 & 0.90 &  0.80 &  0.74 & 0.10 & 0.11 & 0.94 &  0.73 &  0.67 & 0.12 & 0.10 & 0.96 &  0.73 &  0.69 & 0.08 & 0.08 & 0.92 \\
$\rho_{13}$  &  0.26 &  0.20 & 0.28 & 0.22 & 0.99 &  0.52 &  0.50 & 0.13 & 0.14 & 0.94 &  0.42 &  0.38 & 0.20 & 0.16 & 0.98 &  0.44 &  0.37 & 0.20 & 0.16 & 0.98 &  0.65 &  0.61 & 0.11 & 0.11 & 0.96 \\
$\rho_{23}$  &  0.11 &  0.07 & 0.31 & 0.25 & 1.00 &  0.12 &  0.08 & 0.17 & 0.17 & 0.95 &  0.49 &  0.42 & 0.20 & 0.20 & 0.93 &  0.65 &  0.52 & 0.19 & 0.19 & 0.94 &  0.47 &  0.42 & 0.15 & 0.13 & 0.94 \\[1.ex]
\bottomrule
\end{tabular}}
\end{center}
\end{table*}

\begin{table*}[!ht]
\caption{{\bf 1st-order neighbourhood model}. Average values of posterior mean, posterior standard deviation (SD), simulated standard errors (sim) and empirical coverage of the 95\% credible intervals (Cov) for local estimates model parameters based on 100 simulated data sets for Scenario 2.}
\label{tab:Scenario2_k1}
\vspace{-0.5cm}
\renewcommand{\arraystretch}{1.1}
\begin{center}
\resizebox{1.4\textwidth}{!}{
\begin{tabular}{c|crrrr|crrrr|crrrr|crrrr|crrrr}
\toprule
& \multicolumn{5}{c|}{\bf Andaluc\'ia} & \multicolumn{5}{c|}{\bf Arag\'on} & \multicolumn{5}{c|}{\bf Asturias} & \multicolumn{5}{c|}{\bf Cantabria} & \multicolumn{5}{c}{\bf Castilla - La Mancha}\\
& & & & & & & & & & & & & & & & & & & & & \\
Parameter & Value & Mean & SD & Sim & Cov & Value & Mean & SD & Sim & Cov & Value & Mean & SD & Sim & Cov & Value & Mean & SD & Sim & Cov & Value & Mean & SD & Sim & Cov \\
\hline
$\alpha_1$   & -0.20 & -0.20 & 0.01 & 0.02 & 0.86 & -0.20 & -0.20 & 0.03 & 0.04 & 0.79 & -0.20 & -0.19 & 0.03 & 0.07 & 0.67 & -0.20 & -0.19 & 0.04 & 0.07 & 0.78 & -0.20 & -0.21 & 0.02 & 0.02 & 0.87 \\
$\alpha_2$   & -0.10 & -0.09 & 0.02 & 0.02 & 0.90 & -0.10 & -0.10 & 0.03 & 0.04 & 0.91 & -0.10 & -0.09 & 0.04 & 0.07 & 0.70 & -0.10 & -0.08 & 0.05 & 0.07 & 0.83 & -0.10 & -0.11 & 0.02 & 0.02 & 0.93 \\
$\alpha_3$   &  0.10 &  0.10 & 0.02 & 0.02 & 0.93 &  0.10 &  0.10 & 0.04 & 0.04 & 0.97 &  0.10 &  0.10 & 0.05 & 0.06 & 0.93 &  0.10 &  0.09 & 0.06 & 0.07 & 0.90 &  0.10 &  0.09 & 0.03 & 0.03 & 0.90 \\
$\sigma^2_1$ &  0.50 &  0.57 & 0.05 & 0.05 & 0.59 &  0.50 &  0.74 & 0.10 & 0.12 & 0.28 &  0.50 &  0.84 & 0.18 & 0.26 & 0.47 &  0.50 &  0.90 & 0.20 & 0.32 & 0.46 &  0.50 &  0.70 & 0.06 & 0.09 & 0.13 \\
$\sigma^2_2$ &  0.40 &  0.46 & 0.05 & 0.06 & 0.78 &  0.40 &  0.60 & 0.10 & 0.12 & 0.49 &  0.40 &  0.67 & 0.17 & 0.20 & 0.58 &  0.40 &  0.75 & 0.20 & 0.26 & 0.51 &  0.40 &  0.55 & 0.06 & 0.09 & 0.29 \\
$\sigma^2_3$ &  0.30 &  0.35 & 0.05 & 0.06 & 0.82 &  0.30 &  0.46 & 0.12 & 0.14 & 0.65 &  0.30 &  0.52 & 0.17 & 0.16 & 0.80 &  0.30 &  0.62 & 0.21 & 0.25 & 0.65 &  0.30 &  0.42 & 0.07 & 0.09 & 0.52 \\[1.ex]
$\rho_{12}$  &  0.76 &  0.75 & 0.04 & 0.04 & 0.93 &  0.58 &  0.59 & 0.09 & 0.10 & 0.89 &  0.71 &  0.65 & 0.11 & 0.12 & 0.93 &  0.37 &  0.43 & 0.15 & 0.18 & 0.87 &  0.60 &  0.60 & 0.06 & 0.08 & 0.86 \\
$\rho_{13}$  &  0.52 &  0.52 & 0.07 & 0.08 & 0.93 &  0.30 &  0.36 & 0.14 & 0.18 & 0.84 &  0.34 &  0.35 & 0.18 & 0.18 & 0.93 &  0.14 &  0.27 & 0.20 & 0.21 & 0.83 &  0.71 &  0.62 & 0.08 & 0.09 & 0.74 \\
$\rho_{23}$  &  0.47 &  0.48 & 0.08 & 0.08 & 0.91 &  0.37 &  0.42 & 0.15 & 0.15 & 0.92 &  0.51 &  0.45 & 0.17 & 0.18 & 0.93 &  0.69 &  0.62 & 0.16 & 0.15 & 0.97 &  0.38 &  0.39 & 0.10 & 0.13 & 0.84 \\[1.ex]
\hline
& & & & & & & & & & & & & & & & & & & & & \\[-2.ex]
& \multicolumn{5}{c|}{\bf Castilla y Le\'on} & \multicolumn{5}{c|}{\bf Catalu\~na} & \multicolumn{5}{c|}{\bf Comunidad Valenciana} & \multicolumn{5}{c|}{\bf Extremadura} & \multicolumn{5}{c}{\bf Galicia} \\
& & & & & & & & & & & & & & & & & & & & & \\
Parameter & Value & Mean & SD & Sim & Cov & Value & Mean & SD & Sim & Cov & Value & Mean & SD & Sim & Cov & Value & Mean & SD & Sim & Cov & Value & Mean & SD & Sim & Cov \\
\hline
$\alpha_1$   & -0.20 & -0.20 & 0.02 & 0.03 & 0.88 & -0.20 & -0.20 & 0.02 & 0.02 & 0.93 & -0.20 & -0.19 & 0.02 & 0.02 & 0.86 & -0.20 & -0.19 & 0.02 & 0.04 & 0.77 & -0.20 & -0.20 & 0.02 & 0.03 & 0.75 \\
$\alpha_2$   & -0.10 & -0.11 & 0.03 & 0.03 & 0.91 & -0.10 & -0.10 & 0.02 & 0.02 & 0.94 & -0.10 & -0.10 & 0.02 & 0.03 & 0.81 & -0.10 & -0.09 & 0.03 & 0.04 & 0.78 & -0.10 & -0.10 & 0.02 & 0.03 & 0.86 \\
$\alpha_3$   &  0.10 &  0.10 & 0.03 & 0.03 & 0.96 &  0.10 &  0.10 & 0.03 & 0.03 & 0.94 &  0.10 &  0.10 & 0.03 & 0.03 & 0.96 &  0.10 &  0.09 & 0.04 & 0.05 & 0.92 &  0.10 &  0.10 & 0.03 & 0.03 & 0.91 \\
$\sigma^2_1$ &  0.50 &  0.77 & 0.07 & 0.11 & 0.06 &  0.50 &  0.56 & 0.05 & 0.06 & 0.78 &  0.50 &  0.63 & 0.07 & 0.10 & 0.48 &  0.50 &  0.71 & 0.10 & 0.11 & 0.41 &  0.50 &  0.58 & 0.07 & 0.08 & 0.81 \\
$\sigma^2_2$ &  0.40 &  0.63 & 0.07 & 0.10 & 0.12 &  0.40 &  0.45 & 0.05 & 0.06 & 0.81 &  0.40 &  0.52 & 0.06 & 0.08 & 0.53 &  0.40 &  0.57 & 0.10 & 0.11 & 0.51 &  0.40 &  0.45 & 0.06 & 0.07 & 0.84 \\
$\sigma^2_3$ &  0.30 &  0.48 & 0.09 & 0.13 & 0.46 &  0.30 &  0.33 & 0.05 & 0.05 & 0.89 &  0.30 &  0.39 & 0.06 & 0.08 & 0.72 &  0.30 &  0.41 & 0.11 & 0.11 & 0.83 &  0.30 &  0.33 & 0.07 & 0.08 & 0.87 \\[1.ex]
$\rho_{12}$  &  0.60 &  0.63 & 0.06 & 0.07 & 0.83 &  0.54 &  0.54 & 0.06 & 0.07 & 0.88 &  0.72 &  0.71 & 0.05 & 0.06 & 0.88 &  0.30 &  0.40 & 0.10 & 0.11 & 0.80 &  0.61 &  0.63 & 0.06 & 0.07 & 0.89 \\
$\rho_{13}$  &  0.12 &  0.30 & 0.10 & 0.13 & 0.47 &  0.34 &  0.34 & 0.09 & 0.10 & 0.93 &  0.81 &  0.74 & 0.06 & 0.08 & 0.73 &  0.27 &  0.32 & 0.14 & 0.15 & 0.93 &  0.36 &  0.36 & 0.10 & 0.11 & 0.94 \\
$\rho_{23}$  &  0.56 &  0.53 & 0.09 & 0.12 & 0.90 &  0.48 &  0.47 & 0.09 & 0.08 & 0.97 &  0.79 &  0.75 & 0.06 & 0.06 & 0.92 &  0.24 &  0.29 & 0.16 & 0.17 & 0.90 &  0.17 &  0.19 & 0.12 & 0.12 & 0.93 \\[1.ex]
\hline
& & & & & & & & & & & & & & & & & & & & & \\[-2.ex]
& \multicolumn{5}{c|}{\bf La Rioja} & \multicolumn{5}{c}{\bf Madrid} & \multicolumn{5}{c|}{\bf Murcia} & \multicolumn{5}{c|}{\bf Navarra} & \multicolumn{5}{c}{\bf Pa\'is Vasco} \\
& & & & & & & & & & & & & & & & & & & & & \\
Parameter & Value & Mean & SD & Sim & Cov & Value & Mean & SD & Sim & Cov & Value & Mean & SD & Sim & Cov & Value & Mean & SD & Sim & Cov & Value & Mean & SD & Sim & Cov \\
\hline
$\alpha_1$   & -0.20 & -0.23 & 0.05 & 0.08 & 0.79 & -0.20 & -0.20 & 0.03 & 0.06 & 0.70 & -0.20 & -0.22 & 0.03 & 0.09 & 0.53 & -0.20 & -0.21 & 0.04 & 0.05 & 0.84 & -0.20 & -0.20 & 0.03 & 0.04 & 0.84 \\
$\alpha_2$   & -0.10 & -0.13 & 0.06 & 0.10 & 0.76 & -0.10 & -0.11 & 0.04 & 0.05 & 0.87 & -0.10 & -0.12 & 0.03 & 0.08 & 0.58 & -0.10 & -0.11 & 0.04 & 0.06 & 0.83 & -0.10 & -0.10 & 0.03 & 0.04 & 0.88 \\
$\alpha_3$   &  0.10 &  0.08 & 0.08 & 0.10 & 0.84 &  0.10 &  0.09 & 0.05 & 0.06 & 0.93 &  0.10 &  0.09 & 0.05 & 0.08 & 0.71 &  0.10 &  0.08 & 0.06 & 0.06 & 0.93 &  0.10 &  0.10 & 0.04 & 0.05 & 0.89 \\
$\sigma^2_1$ &  0.50 &  1.12 & 0.24 & 0.29 & 0.19 &  0.50 &  0.87 & 0.13 & 0.19 & 0.16 &  0.50 &  1.06 & 0.23 & 0.25 & 0.14 &  0.50 &  0.85 & 0.15 & 0.20 & 0.29 &  0.50 &  0.83 & 0.12 & 0.22 & 0.27 \\
$\sigma^2_2$ &  0.40 &  0.90 & 0.24 & 0.33 & 0.36 &  0.40 &  0.67 & 0.12 & 0.16 & 0.37 &  0.40 &  0.86 & 0.20 & 0.26 & 0.29 &  0.40 &  0.64 & 0.14 & 0.17 & 0.52 &  0.40 &  0.64 & 0.11 & 0.17 & 0.47 \\
$\sigma^2_3$ &  0.30 &  0.81 & 0.29 & 0.33 & 0.48 &  0.30 &  0.51 & 0.12 & 0.16 & 0.53 &  0.30 &  0.65 & 0.19 & 0.23 & 0.45 &  0.30 &  0.56 & 0.17 & 0.19 & 0.62 &  0.30 &  0.50 & 0.12 & 0.15 & 0.56 \\[1.ex]
$\rho_{12}$  &  0.65 &  0.61 & 0.13 & 0.16 & 0.92 &  0.66 &  0.66 & 0.08 & 0.11 & 0.88 &  0.80 &  0.73 & 0.08 & 0.11 & 0.86 &  0.73 &  0.71 & 0.09 & 0.10 & 0.90 &  0.73 &  0.71 & 0.07 & 0.10 & 0.83 \\
$\rho_{13}$  &  0.26 &  0.39 & 0.19 & 0.20 & 0.81 &  0.52 &  0.49 & 0.12 & 0.15 & 0.90 &  0.42 &  0.45 & 0.15 & 0.18 & 0.85 &  0.44 &  0.46 & 0.15 & 0.17 & 0.89 &  0.65 &  0.63 & 0.10 & 0.13 & 0.88 \\
$\rho_{23}$  &  0.11 &  0.27 & 0.21 & 0.26 & 0.83 &  0.12 &  0.26 & 0.14 & 0.18 & 0.82 &  0.49 &  0.54 & 0.14 & 0.16 & 0.90 &  0.65 &  0.50 & 0.15 & 0.19 & 0.81 &  0.47 &  0.47 & 0.13 & 0.15 & 0.90 \\[1.ex]
\bottomrule
\end{tabular}}
\end{center}
\end{table*}

\begin{table*}[!ht]
\caption{{\bf 2nd-order neighbourhood model}. Average values of posterior mean, posterior standard deviation (SD), simulated standard errors (sim) and empirical coverage of the 95\% credible intervals (Cov) for local estimates model parameters based on 100 simulated data sets for Scenario 2.}
\label{tab:Scenario2_k2}
\vspace{-0.5cm}
\renewcommand{\arraystretch}{1.1}
\begin{center}
\resizebox{1.4\textwidth}{!}{
\begin{tabular}{c|crrrr|crrrr|crrrr|crrrr|crrrr}
\toprule
& \multicolumn{5}{c|}{\bf Andaluc\'ia} & \multicolumn{5}{c|}{\bf Arag\'on} & \multicolumn{5}{c|}{\bf Asturias} & \multicolumn{5}{c|}{\bf Cantabria} & \multicolumn{5}{c}{\bf Castilla - La Mancha}\\
& & & & & & & & & & & & & & & & & & & & & \\
Parameter & Value & Mean & SD & Sim & Cov & Value & Mean & SD & Sim & Cov & Value & Mean & SD & Sim & Cov & Value & Mean & SD & Sim & Cov & Value & Mean & SD & Sim & Cov \\
\hline
$\alpha_1$   & -0.20 & -0.20 & 0.01 & 0.03 & 0.63 & -0.20 & -0.19 & 0.02 & 0.06 & 0.52 & -0.20 & -0.19 & 0.03 & 0.10 & 0.52 & -0.20 & -0.19 & 0.04 & 0.09 & 0.60 & -0.20 & -0.20 & 0.02 & 0.02 & 0.80 \\
$\alpha_2$   & -0.10 & -0.09 & 0.01 & 0.03 & 0.71 & -0.10 & -0.09 & 0.03 & 0.05 & 0.68 & -0.10 & -0.09 & 0.04 & 0.09 & 0.58 & -0.10 & -0.08 & 0.04 & 0.09 & 0.64 & -0.10 & -0.11 & 0.02 & 0.02 & 0.89 \\
$\alpha_3$   &  0.10 &  0.10 & 0.02 & 0.03 & 0.79 &  0.10 &  0.10 & 0.04 & 0.04 & 0.88 &  0.10 &  0.11 & 0.05 & 0.09 & 0.70 &  0.10 &  0.10 & 0.06 & 0.09 & 0.79 &  0.10 &  0.10 & 0.02 & 0.03 & 0.93 \\
$\sigma^2_1$ &  0.50 &  0.57 & 0.05 & 0.05 & 0.63 &  0.50 &  0.72 & 0.08 & 0.11 & 0.24 &  0.50 &  0.80 & 0.16 & 0.23 & 0.49 &  0.50 &  0.86 & 0.17 & 0.27 & 0.48 &  0.50 &  0.71 & 0.05 & 0.08 & 0.06 \\
$\sigma^2_2$ &  0.40 &  0.46 & 0.05 & 0.06 & 0.78 &  0.40 &  0.58 & 0.09 & 0.11 & 0.43 &  0.40 &  0.64 & 0.15 & 0.18 & 0.59 &  0.40 &  0.72 & 0.17 & 0.23 & 0.40 &  0.40 &  0.57 & 0.06 & 0.09 & 0.21 \\
$\sigma^2_3$ &  0.30 &  0.35 & 0.05 & 0.06 & 0.81 &  0.30 &  0.45 & 0.10 & 0.13 & 0.60 &  0.30 &  0.50 & 0.15 & 0.14 & 0.75 &  0.30 &  0.57 & 0.18 & 0.19 & 0.65 &  0.30 &  0.44 & 0.07 & 0.08 & 0.36 \\[1.ex]
$\rho_{12}$  &  0.76 &  0.74 & 0.04 & 0.04 & 0.94 &  0.58 &  0.60 & 0.08 & 0.09 & 0.88 &  0.71 &  0.63 & 0.10 & 0.12 & 0.91 &  0.37 &  0.48 & 0.13 & 0.16 & 0.80 &  0.60 &  0.59 & 0.05 & 0.07 & 0.86 \\
$\rho_{13}$  &  0.52 &  0.49 & 0.07 & 0.07 & 0.91 &  0.30 &  0.37 & 0.12 & 0.15 & 0.83 &  0.34 &  0.35 & 0.16 & 0.17 & 0.92 &  0.14 &  0.34 & 0.17 & 0.19 & 0.75 &  0.71 &  0.57 & 0.07 & 0.09 & 0.48 \\
$\rho_{23}$  &  0.47 &  0.47 & 0.07 & 0.08 & 0.90 &  0.37 &  0.44 & 0.13 & 0.13 & 0.86 &  0.51 &  0.45 & 0.16 & 0.17 & 0.92 &  0.69 &  0.59 & 0.15 & 0.15 & 0.94 &  0.38 &  0.40 & 0.09 & 0.13 & 0.71 \\[1.ex]
\hline
& & & & & & & & & & & & & & & & & & & & & \\[-2.ex]
& \multicolumn{5}{c|}{\bf Castilla y Le\'on} & \multicolumn{5}{c|}{\bf Catalu\~na} & \multicolumn{5}{c|}{\bf Comunidad Valenciana} & \multicolumn{5}{c|}{\bf Extremadura} & \multicolumn{5}{c}{\bf Galicia} \\
& & & & & & & & & & & & & & & & & & & & & \\
Parameter & Value & Mean & SD & Sim & Cov & Value & Mean & SD & Sim & Cov & Value & Mean & SD & Sim & Cov & Value & Mean & SD & Sim & Cov & Value & Mean & SD & Sim & Cov \\
\hline
$\alpha_1$   & -0.20 & -0.20 & 0.02 & 0.03 & 0.75 & -0.20 & -0.20 & 0.02 & 0.02 & 0.83 & -0.20 & -0.19 & 0.02 & 0.04 & 0.71 & -0.20 & -0.19 & 0.02 & 0.07 & 0.39 & -0.20 & -0.20 & 0.02 & 0.05 & 0.46 \\
$\alpha_2$   & -0.10 & -0.10 & 0.02 & 0.03 & 0.78 & -0.10 & -0.10 & 0.02 & 0.02 & 0.88 & -0.10 & -0.09 & 0.02 & 0.04 & 0.70 & -0.10 & -0.09 & 0.03 & 0.07 & 0.54 & -0.10 & -0.10 & 0.02 & 0.05 & 0.55 \\
$\alpha_3$   &  0.10 &  0.10 & 0.03 & 0.04 & 0.88 &  0.10 &  0.10 & 0.03 & 0.03 & 0.91 &  0.10 &  0.10 & 0.03 & 0.03 & 0.86 &  0.10 &  0.10 & 0.03 & 0.06 & 0.74 &  0.10 &  0.09 & 0.03 & 0.05 & 0.74 \\
$\sigma^2_1$ &  0.50 &  0.74 & 0.06 & 0.09 & 0.03 &  0.50 &  0.56 & 0.05 & 0.06 & 0.76 &  0.50 &  0.63 & 0.06 & 0.10 & 0.47 &  0.50 &  0.70 & 0.09 & 0.11 & 0.33 &  0.50 &  0.58 & 0.07 & 0.08 & 0.76 \\
$\sigma^2_2$ &  0.40 &  0.60 & 0.06 & 0.09 & 0.13 &  0.40 &  0.45 & 0.05 & 0.06 & 0.81 &  0.40 &  0.52 & 0.06 & 0.08 & 0.49 &  0.40 &  0.57 & 0.09 & 0.11 & 0.49 &  0.40 &  0.46 & 0.06 & 0.07 & 0.82 \\
$\sigma^2_3$ &  0.30 &  0.46 & 0.07 & 0.11 & 0.40 &  0.30 &  0.34 & 0.05 & 0.05 & 0.88 &  0.30 &  0.39 & 0.06 & 0.08 & 0.67 &  0.30 &  0.42 & 0.10 & 0.10 & 0.81 &  0.30 &  0.34 & 0.06 & 0.08 & 0.91 \\[1.ex]
$\rho_{12}$  &  0.60 &  0.63 & 0.05 & 0.07 & 0.79 &  0.54 &  0.54 & 0.06 & 0.07 & 0.90 &  0.72 &  0.71 & 0.05 & 0.06 & 0.87 &  0.30 &  0.43 & 0.09 & 0.11 & 0.66 &  0.61 &  0.63 & 0.06 & 0.07 & 0.88 \\
$\rho_{13}$  &  0.12 &  0.35 & 0.08 & 0.12 & 0.29 &  0.34 &  0.35 & 0.09 & 0.10 & 0.91 &  0.81 &  0.69 & 0.06 & 0.08 & 0.48 &  0.27 &  0.33 & 0.13 & 0.15 & 0.89 &  0.36 &  0.37 & 0.10 & 0.11 & 0.92 \\
$\rho_{23}$  &  0.56 &  0.50 & 0.08 & 0.11 & 0.85 &  0.48 &  0.47 & 0.08 & 0.08 & 0.97 &  0.79 &  0.72 & 0.06 & 0.07 & 0.81 &  0.24 &  0.33 & 0.14 & 0.16 & 0.85 &  0.17 &  0.20 & 0.11 & 0.12 & 0.90 \\[1.ex]
\hline
& & & & & & & & & & & & & & & & & & & & & \\[-2.ex]
& \multicolumn{5}{c|}{\bf La Rioja} & \multicolumn{5}{c}{\bf Madrid} & \multicolumn{5}{c|}{\bf Murcia} & \multicolumn{5}{c|}{\bf Navarra} & \multicolumn{5}{c}{\bf Pa\'is Vasco} \\
& & & & & & & & & & & & & & & & & & & & & \\
Parameter & Value & Mean & SD & Sim & Cov & Value & Mean & SD & Sim & Cov & Value & Mean & SD & Sim & Cov & Value & Mean & SD & Sim & Cov & Value & Mean & SD & Sim & Cov \\
\hline
$\alpha_1$   & -0.20 & -0.23 & 0.04 & 0.10 & 0.60 & -0.20 & -0.21 & 0.03 & 0.09 & 0.53 & -0.20 & -0.24 & 0.02 & 0.17 & 0.21 & -0.20 & -0.21 & 0.03 & 0.07 & 0.64 & -0.20 & -0.20 & 0.03 & 0.06 & 0.64 \\
$\alpha_2$   & -0.10 & -0.12 & 0.05 & 0.11 & 0.55 & -0.10 & -0.11 & 0.04 & 0.07 & 0.72 & -0.10 & -0.12 & 0.03 & 0.15 & 0.34 & -0.10 & -0.11 & 0.04 & 0.07 & 0.72 & -0.10 & -0.11 & 0.03 & 0.06 & 0.65 \\
$\alpha_3$   &  0.10 &  0.07 & 0.07 & 0.11 & 0.75 &  0.10 &  0.09 & 0.05 & 0.07 & 0.79 &  0.10 &  0.08 & 0.04 & 0.13 & 0.39 &  0.10 &  0.09 & 0.05 & 0.07 & 0.80 &  0.10 &  0.10 & 0.04 & 0.06 & 0.78 \\
$\sigma^2_1$ &  0.50 &  1.11 & 0.20 & 0.27 & 0.08 &  0.50 &  0.81 & 0.11 & 0.15 & 0.17 &  0.50 &  0.85 & 0.15 & 0.18 & 0.25 &  0.50 &  0.80 & 0.12 & 0.16 & 0.21 &  0.50 &  0.82 & 0.11 & 0.20 & 0.25 \\
$\sigma^2_2$ &  0.40 &  0.89 & 0.20 & 0.27 & 0.26 &  0.40 &  0.64 & 0.10 & 0.14 & 0.36 &  0.40 &  0.70 & 0.14 & 0.18 & 0.36 &  0.40 &  0.60 & 0.11 & 0.14 & 0.51 &  0.40 &  0.64 & 0.11 & 0.17 & 0.40 \\
$\sigma^2_3$ &  0.30 &  0.82 & 0.25 & 0.33 & 0.28 &  0.30 &  0.50 & 0.11 & 0.14 & 0.53 &  0.30 &  0.55 & 0.14 & 0.18 & 0.49 &  0.30 &  0.52 & 0.13 & 0.15 & 0.56 &  0.30 &  0.50 & 0.11 & 0.15 & 0.54 \\[1.ex]
$\rho_{12}$  &  0.65 &  0.62 & 0.11 & 0.15 & 0.86 &  0.66 &  0.63 & 0.07 & 0.10 & 0.88 &  0.80 &  0.73 & 0.07 & 0.10 & 0.78 &  0.73 &  0.71 & 0.08 & 0.09 & 0.90 &  0.73 &  0.70 & 0.07 & 0.10 & 0.86 \\
$\rho_{13}$  &  0.26 &  0.42 & 0.16 & 0.19 & 0.78 &  0.52 &  0.45 & 0.11 & 0.14 & 0.81 &  0.42 &  0.49 & 0.12 & 0.14 & 0.85 &  0.44 &  0.48 & 0.13 & 0.15 & 0.88 &  0.65 &  0.61 & 0.10 & 0.13 & 0.87 \\
$\rho_{23}$  &  0.11 &  0.32 & 0.18 & 0.22 & 0.75 &  0.12 &  0.28 & 0.13 & 0.16 & 0.71 &  0.49 &  0.56 & 0.12 & 0.14 & 0.85 &  0.65 &  0.46 & 0.14 & 0.17 & 0.72 &  0.47 &  0.46 & 0.12 & 0.15 & 0.89 \\[1.ex]
\bottomrule
\end{tabular}}
\end{center}
\end{table*}

\end{landscape}


\end{document}